# Mapping the Electronic Structure Origins of Surface- and Chemistry-Dependent Doping Trends in III-V Quantum Dots

Michael G. Taylor[1] and Heather J. Kulik[1,*]

[1]*Department of Chemical Engineering, Massachusetts Institute of Technology, Cambridge, MA 02139*

ABSTRACT: Modifying the optoelectronic properties of nanostructured materials through introduction of dopant atoms has attracted intense interest. Nevertheless, the approaches employed are often trial and error, preventing rational design. We demonstrate the power of large-scale electronic structure calculations with density functional theory (DFT) to build an atlas of preferential dopant sites for a range of M(II) and M(III) dopants in the representative III-V InP magic sized cluster (MSC). We quantify the thermodynamic favorability of dopants, which we identify to be both specific to the sites within the MSC (i.e., interior vs surface) and to the nature of the dopant atom (i.e., smaller Ga(III) vs larger Y(III) or Sc(III)). These observations motivate development of maps of the most and least favorable doping sites, which are consistent with some known experimental expectations but also yield unexpected observations. For isovalent doping (i.e., Y(III)/Sc(III) or Ga(III), we observed stronger sensitivity of the predicted energetics to the type of ligand orientation on the surface than to the dopant type, but divergent behavior is observed for whether interior doping is favorable. For charge balancing with M(II) (i.e., Zn or Cd) dopants, we show that the type of ligand removed during the doping reaction is critical. We show that limited cooperativity with dopants up to moderate concentrations occurs, indicating rapid single-dopant estimations of favorability from DFT can efficiently guide rational design. Our work emphasizes the strong importance of ligand chemistry and surface heterogeneity in determining paths to favorable doping in quantum dots, an observation that will be general to other III-V and II-VI quantum dot systems generally synthesized with carboxylate ligands.

# 1. Introduction.

Colloidal quantum dots (QDs) have unique size-dependent electronic and optical properties[1-2] that make them ideal for applications in photovoltaics,[3-5] light-emitting diodes,[6-7] and biological imaging.[8-9] The III-V InP QDs have emerged as a less toxic[10] alternative[11-12] with a broader emission range[13-14] than the well studied II-VI CdSe QDs that have tunable sizes[15-16] and exhibit emission throughout the visible range.[17] Adoption of InP QDs has been limited by difficulties[11] in obtaining optimally narrow QD size distributions and high photoluminescent quantum yields, despite ongoing experimental efforts to understand the QD formation mechanism[18-23] and tune the growth process.[24-25] The typical InP synthesis recipe requires high temperatures, employs indium carboxylates[26-28] that necessitate reactive phosphorus precursors, and long reaction times, making control of size distributions challenging.[29] One possibility for improving InP QDs is through the judicious doping of materials either at low concentrations or to form core-shell QDs with both isovalent and heterovalent dopants.

M(II) transition metal dopants (e.g. Cd[30-32], Mn[33-34], Cu[35-37] and Zn[31, 35-36]) are commonly incorporated into the InP nanoclusters and used in shells of InP QDs. In particular, both Zn(II) and Cd(II) dopants are known to enhance quantum yield (QY) in InP QDs[31] and have been successfully cation exchanged into the MSC.[30, 35] Additionally, Zn(II) is commonly incorporated in ZnSe/ZnS shells on InP core-shell QDs.[38] Cd dopants are less widely investigated in InP shells due to their toxicity, but are still of interest for capturing site- and concentration-specific enhancement of QY.[31] Further, dynamic nuclear polarization surface-enhanced NMR spectroscopy was applied to specifically determine possible locations of Cd ions and different surface species in Cd-doped InP QDs, which enables comparison with computational predictions of Cd doping sites.[32] We analyze the role of Cd(II) and Zn(II) site-specific doping in the MSC to interrogate positional effects in the MSC while remaining close to experimental observations. However, a clear combinatorial challenge arises in identifying how to tune precursor chemistry during synthesis and during doping.

Despite the importance of doping as a path to improve InP QD properties, this challenge in possible synthesis recipes requires that we improve knowledge *a priori* regarding how and where dopants will influence the local electronic and geometric structure. A promising avenue to increase understanding of how doping will improve InP properties is through well-controlled experimental and computational study of model systems. Specifically, InP magic size clusters (MSCs)[19, 39] around 1.1[40] to 1.3[40-41] nm in size have been observed and structurally characterized as intermediates during InP growth, with their size[40] and stability[19] controlled by In precursor chemistry. InP MSCs have been used as seeds for controlled QD growth[19] by decoupling the nucleation and growth stages[21]. In addition to aiding understanding III-V InP[19, 41-42] and InAs[43] growth mechanisms, controlled growth of MSCs has also been observed to be beneficial in II-VI materials (e.g., CdS[44], CdTe[45], and ternary II-VI alloys[46]). First-principles simulations have been a valuable complement to these experimental studies in understanding the growth mechanism[47-48] of II-VI and IV-VI QDs. We have computationally characterized the surface of both early stage (i.e., the smallest sized) clusters[49] computationally "synthesized" with *ab initio* molecular dynamics (AIMD) as well as MSCs and made predictions on structures of larger QDs. Although properties of amorphous[50], bulk[51-53], and quantum-confined InP[54-58] had been studied before, our study allowed us to develop a unified electronic structure interpretation of surface reactivity on early-stage[49] and magic sized[41] clusters.

In these computational studies, we observed a strong influence of carboxylate binding mode in determining surface reactivity. Carboxylate precursors are widely used for both In[19] and cationic dopants[30, 35], especially with the long-chain myristic acid[35-36] (MA). Our observations predicted the outcomes of subsequent experiments[59] that binding mode strongly influences growth, and other experiments[28, 60] have shown comparable binding-mode dependence at larger QD surfaces. Given these strong trends with ligand binding mode, we can anticipate that electronic structure calculations of doping reaction steps will depend on the mode of carboxylate ligand binding. Given the good correspondence of the MSC as a bridge between early stage and QD clusters, it also represents an excellent platform to identify trends in how dopant identity influences doping favorability and alters the electronic properties of InP QDs. The rest of this article is outlined as follows. In section 2, we provide an overview of the models and dopants studied. In section 3, we present the computational details of the work. In section 4, we present results and discussion of both isovalent M(III) and M(II) dopants into a magic sized cluster. Finally, in section 5, we present our conclusions.

**2. Structures of InP Clusters.**

Before evaluating the structural-sensitivity of doping energetics, we review the structural features of the crystallographically resolved InP magic sized cluster (MSC). The 1.3 nm diameter MSC[41] contains a near-1:1 $[In_{21}P_{20}]^{3+}$ core with each P coordinated to 4 In in pseudotetrahedral conformations surrounded by 16 excess surface In (Figure 1). Because the In sites on the MSC range from those that are near bulk-like to those that are undercoordinated at the surface, it has been exploited as an ideal bridge between early stage clusters to full-scale QDs and bulk materials.[61] While the coordination numbers (CNs) of In and P and P-In-P angles of the core InP mirror that of bulk P in zincblende InP[62], other structural features (i.e., torsional In and P angles) deviate from those in the bulk.[41] If we categorize each In in the MSC by the number of In-P bonds formed with P, only a minority are solely-coordinated by P (i.e., seven form four In-P bonds and eight that form three In-P bonds), whereas more are coordinated by carboxylates and thus have lower coordination by P (i.e., six In with two In-P bonds and 16 with one In-P bond, Figure 1 and Supporting Information Table S1). Since the In-P coordination environment coincides with the location of In in the MSC, we designate In atoms that form three bonds with P *interior* sites and those that form four bonds as *bulk* sites (Figure 1). Since the In atoms that coordinate even fewer P appear on the surface, we designate those that form one or two bonds with P as *surface* sites (Figure 1).

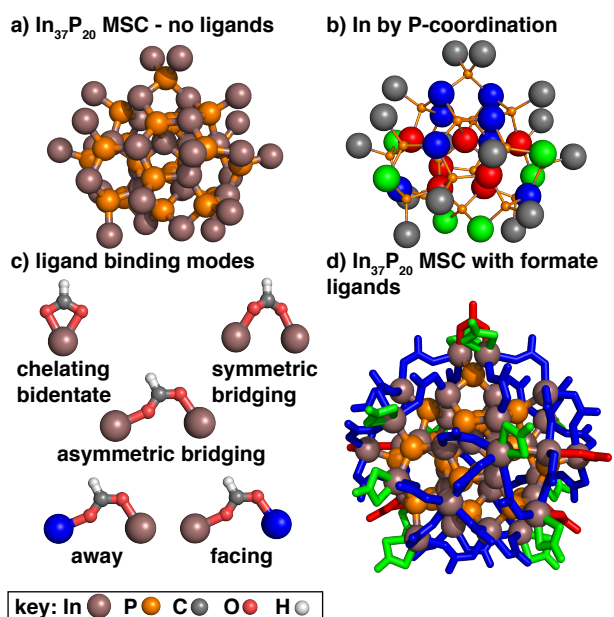

**Figure 1.** Structure and nomenclature for sites on the $In_{37}P_{20}$ MSC used through this work: (a) MSC without ligands; (b) MSC without ligands with In sites colored by number of coordinating P-atoms (gray = 1, green = 2, blue =3, and red = 4); (c) carboxylate ligand binding modes shown with formate; (d) MSC with formate ligands shown as sticks colored by ligand binding mode (green = chelating bidentate, blue = asymmetric bridging, and red = symmetric bridging).

Notably, within surface sites differences are anticipated due to distinctions in carboxylate ligand binding modes that vary across the MSC. Carboxylate ligands are known[28, 41, 63] to bind in different configurations at the surface of InP materials ranging from chelating bidentate (CB) modes bound to a single In center to symmetric modes bridging (SB) between In centers (Figure 1). Although a range of terminologies[28, 41, 49, 61, 63-64] have been adopted, we designate the intermediate modes with partial CB and SB character as asymmetric bridging (AB) modes that can either be facing (ABF) or away (ABA) from an In atom, (Figure 1 and Supporting Information Table S2). Monodentate carboxylate ligand binding modes are not observed in the structurally-characterized surface of the MSC[41], likely due to the high energy for their formation[49, 61]. Analysis of the surface of the MSC reveals symmetric and distributed CB and SB sites interspersed between AB sites (Figure 1 and Supporting Information Table S3). Given the ligand binding mode-dependence of surface sites[61], the binding mode can be expected to influence doping energetics and properties as well.

There are 12 CB ligands[61] in the MSC that bind a single In site and 39 SB or AB ligands that bridge multiple surface sites. Thus, for dopant with M(II) (i.e., Cd or Zn), there are 90 distinct ways in which the formate ligands can be removed on the surface (see Sec. 3). We evaluate all of these sites (i.e., for Cd) and a subset for Zn (Supporting Information Figure S1). In all cases, we truncate carboxylate ligands to formate after confirming the effect on dopant energetics is modest (e.g., for Zn(II) dopants, see Sec. 3 and Supporting Information Figure S2 and Table S4).

## 3. Computational Details.

*Electronic structure calculations.* All electronic structure calculations of the MSC were carried out using density functional theory (DFT) with a developer version of TeraChem v1.9.[65] The DFT geometry optimizations employed the hybrid B3LYP[66-68] exchange-correlation (xc) functional following the default definition in TeraChem that uses the VWN1-RPA form for the local density approximation of the correlation.[69] Empirical dispersion correction (D3) was added[70] with the default Becke-Johnson damping[71]. All calculations employed the LACVP* composite basis set, which corresponds to an LANL2DZ[72] effective core potential for heavy atoms (i.e., Cd, Ga, In, Sc, Y, and Zn) and the 6-31G* basis for all other atoms. These functional and basis set choices were made to enable large-scale optimization of hundreds of dopant/site combinations and were confirmed to have a limited effect on relative energetics for each dopant addition on either the MSC or a representative (i.e., early-stage cluster) model from prior simulations[49] (Supporting Information Figures S3–S4 and Tables S5–S6).

Given the large system sizes (ca. 200-400 atoms) and in accordance with our prior work[61], all dopant reaction energetics were computed as gas phase, electronic energies. Geometry optimizations were carried out using the DL-FIND[73] interface to TeraChem implementing with L-BFGS in Cartesian coordinates. Default thresholds of $4.5 \times 10^{-4}$ hartree/bohr for the maximum gradient and $1.0 \times 10^{-6}$ for the change in energy were used for all optimizations. The choice of gas phase for the optimizations was motivated by the low dielectric solvents (i.e., toluene, $\varepsilon = 2.38$)[74] commonly used in experimental synthesis of InP QDs.[19, 30, 41] We confirmed a limited effect of including solvent by carrying out tests that incorporated an implicit conductor-like polarizable continuum model (C-PCM)[75-76] implemented in TeraChem[77-78] on Zn-doping energetics (Supporting Information Figure S5 and Table S7).

*Model generation.* The initial MSC computational models for geometry optimization were generated by reducing phenylacetate in the X-ray crystal structure of $In_{37}P_{20}(O_2CCH_2Ph)_{51}$[41] to formate, as in prior work.[61] Here, formate ligands on the MSC were generated by replacing the methyl groups and downstream atoms with a single hydrogen atom along the C-C bond vector and at an initial distance of 1.10 Å (i.e., the C-H bond length in formate). Representative initial In formate precursor geometries were obtained from prior work.[49] For isovalent dopants, (M(III) = Ga, Y, or Sc), initial reference precursors were generated by replacing In with the dopant. Reference precursors for dopants with +2 formal charge (M(II) = Zn or Cd) were generated by coordinating the metals with only two formate ligands. Additional precursor structures were generated by molSimplify[79] with all metals coordinated by acetate, phenylacetate, and butyrate ligands for comparison. For isovalent doped MSCs, the same procedure of replacing each In *surface* site was carried out, whereas for M(II) dopants a formate ligand was removed. Structural property evaluations, including the presence of bonds based on geometric cutoffs (i.e., 1.15x the sum of covalent radii[80]) were evaluated using extensions to molSimplify.[79, 81-82] As surface In could be coordinated to two formate ligands, each possibility for ligand removal was evaluated. All structures were subsequently geometry optimized.

Higher dopant concentrations were studied for Ga, Cd, and Zn dopants in the MSC by adding up to 15 dopant atoms in increments of two in a stepwise fashion. Three initial configurations for each metal were selected based on the energetics of the single-dopant sites: (i) the predicted lowest, (ii) the second lowest combinations of single dopant energies, and (iii) a randomly selected combination of single dopants as long as it was predicted to be exothermic based on the sum of the individual dopant energetics. Of these three configurations, the one with the lowest energy after geometry relaxation was selected for subsequent calculations at higher dopant concentrations. Both initial and final structures for all species are provided in the Supporting Information .zip file.

**4. Results and Discussion.**

**4a. Trends in Size with Isovalent M(III) Doping.**

We first quantify the favorability of single-site doping across In(III) MSC surface sites by isovalent Ga(III). There have been experimental reports of Ga(III) doping of the InP MSC[35] and larger InP QDs.[83-87] It has been suggested that Ga(III) could be doping at the surface[35, 83, 85] or mixed[87] with some degree of alloying depending on conditions. At the least-P-coordinated *surface* MSC sites (i.e., In coordinated to one P, P = 1), Ga(III) doping energetics are near-thermoneutral on average (3.0 kcal/mol) with a small standard deviation (std. dev. = 2.1 kcal/mol, Figure 2 and Supporting Information Tables S1 and S6 and Figure S6). The most favorable P = 1 *surface* site for Ga doping is on an edge of the MSC where the Ga dopant is coordinated by only symmetric or asymmetric bridging ligands (Figure 2 and Supporting Information Table S3). The least favorable P = 1 *surface* site is nearly 8 kcal/mol less favorable is directly adjacent to the most favorable P = 1 *surface* site but is coordinated by a CB ligand (Supporting Information Table S3). In comparison to the P=1 *surface* sites, P = 2 *surface* sites are markedly less favorable (7.5 ± 1.6 kcal/mol) overall (Figure 2 and Supporting Information Table S3). These trends suggest Ga(III) *surface* site preference is most strongly influenced by the coordinating-ligand orientation around the dopant site rather than the MSC core structure.

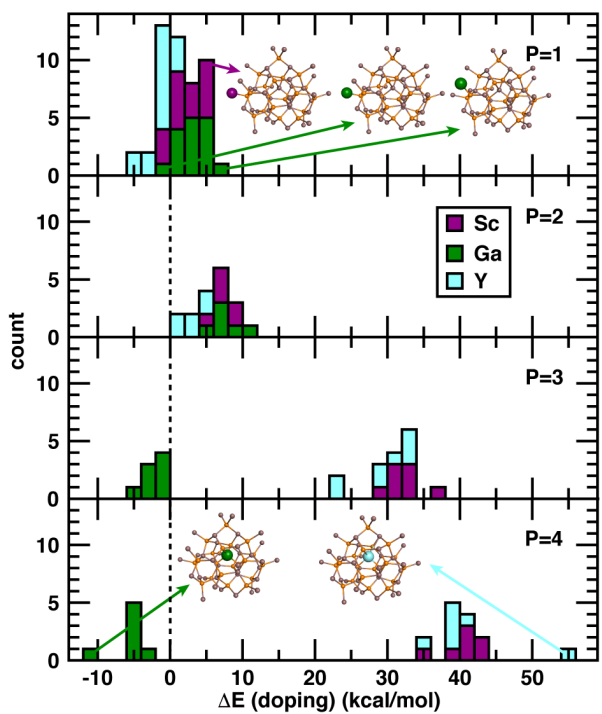

**Figure 2.** Stacked histograms of ΔE (doping) for M(III) dopants (Sc in purple, Ga in dark green, and Y in cyan) into the InP MSC grouped by increasing P-coordination (top to bottom). Inset images are representative relaxed structures of the doped MSC with dopant locations indicated

by an enlarged sphere colored as indicated in the inset legend. The bin width is 2 kcal/mol, and histograms have not been normalized. A black dashed line indicates ΔE (doping)=0 for all graphs.

Since our calculations employed a simplified model (i.e., formate) of carboxylate ligands, we investigated if longer-chain carboxylate ligands more widely used experimentally would influence observed Ga(III) surface site preferences. We did not expect a significant effect because we previously observed unchanged trends in rigid ligand dissociation energies between formate and larger acetate or phenylacetate ligands on the InP MSC.[61] Where modest differences had been observed, these deviations corresponded to cases where inter-ligand hydrogen bonding was altered by the difference in chain length of the carboxylate. As in prior work, exchanging a CB formate on a Ga(III) precursor compound with acetate, butyrate, or phenylacetate[41] shifts ligand binding energies by at most 2.6 kcal/mol (Supporting Information Figure S7). This small difference compared to the range of energetics in the Ga(III) doping at P = 1 versus P = 2 *surface* sites suggests trends in doping site preferences should be preserved with alternate carboxylate ligands (Supporting Information Table S8).

Experimental efforts to dope Ga(III) into a MSC with long-chain myristic acid (MA) carboxylate ligands via introduction of Ga(OAc)$_3$ precursors was observed to lead to formation of Ga-P bonds detected via $^{31}$P NMR.[35] This led researchers to propose that Ga(III) surface doping[35] could be favorable and driven by surface strain reduction on the MSC due to the smaller covalent radius of Ga(III) (i.e., 1.22 Å) in comparison to In(III) (i.e., 1.42 Å). Contrary to this hypothesis, we observe that the most favorable sites for single Ga(III) dopant addition are in *interior* (P=3) and *bulk* (P=4) sites (Figure 2 and Supporting Information Figure S6 and Table S9). Indeed, the most favorable site for Ga(III) addition (-10.6 kcal/mol) is at the center of the MSC (Figure 2 and Supporting Information Figure S6 and Table S9). This favorable interior

Ga(III) doping is also unintuitive since heuristics such as hard-soft acid base (HSAB) would lead to predictions that the harder Ga(III) cation would favor harder, partial carboxylate coordination over the softer P-coordinating *bulk* site (Supporting Information Table S10).[88-89] Nevertheless, this discrepancy can be readily rationalized from observations in our calculations. Despite the smaller covalent radius of Ga(III) relative to In(III), M(III)-P bond lengths at the *interior* P = 3 and *bulk* P = 4 sites are relatively unchanged upon addition of a dopant (Supporting Information Table S11 and Figures S8–S9). Given the limited rearrangement of the MSC, we thus attribute the thermodynamic favorability of interior Ga(III) doping to higher favorability of Ga-P bond formation in comparison to Ga-O bonds. The Ga doping at this lowest energy site also increases the highest occupied molecular orbital (HOMO) to lowest occupied molecular orbital (LUMO) gap (i.e. the fundamental gap) of the Ga-doped MSC by a factor of 1.03×. This increase corresponds to a slight blue shift in agreement with experimental observations[83, 87] of Ga-doping in larger InP QDs (Supporting Information Figure S10 and Table S1). While these calculated preferences for interior Ga(III) doping contrast with expectations, they thus also align with experimentally measured spectral shifts from Ga(III) doping in InP QDs.[83, 87]

To investigate if these trends are sensitive to dopant size, we also evaluated energetics for isovalent Sc(III) or Y(III) doping. The even larger covalent radius of Sc(III) (1.70 Å) in comparison to In(III), would lead to expectations[90] that Sc(III) should exhibit a preference for the surface of the MSC (Supporting Information Table S11). Unlike Ga(III), Sc(III) indeed strongly prefers *surface* site doping, as the highest energy *bulk* P = 4 dopant is unfavorable by nearly 40 kcal/mol with respect to the least favorable surface P =1 site (i.e., 43.5 kcal/mol versus 5.9 kcal/mol, Figure 2). As expected, the even larger Y(III) (1.90 Å) prefers MSC *surface* doping sites even more, with the highest energy *bulk* P = 4 site more than 50 kcal/mol less favorable than the highest energy *surface* P = 1 site (i.e., 54.1 kcal/mol versus 1.2 kcal/mol). This trend can be rationalized based on our earlier observation with Ga(III) doping that interior and bulk site M(III)-P bond lengths do not change as significantly as expected even after the dopant is introduced. As a result, relative to the sum of covalent radii, the largest Sc(III) and Y(III) dopants in interior P = 3 or bulk P = 4 sites have significantly compressed M(III)-P bonds (Supporting Information Figure S8). In the case of Sc(III) and Y(III), *interior* P = 3 or *bulk* P = 4 site doping does induce changes in adjacent In-P bonds but not in a way that leads to overall favorable doping energetics (Supporting Information Figure S9). Thus, while interior sites were favorable for the smaller Ga(III) dopant, the mismatch in size of the larger dopants means the InP MSC cannot rearrange favorably around them.

Despite differences in Ga(III) and Sc(III) or Y(III) size, we still expect ligand binding configuration to play a significant role in relative site favorability for each dopant (i.e., Sc(III) or Y(III)) on the MSC surface (Figure 2). Qualitatively, Y(III) and Sc(III) dopants have similar (i.e., within 1 kcal/mol) average energetics to Ga(III) at both *surface* (i.e., P = 1 and P = 2) sites (Supporting Information Table S12). Despite this qualitative similarity, analysis of specific dopant sites reveals that the least-favorable MSC *surface* site for Sc(III) and Y(III) is the most favored *surface* site for Ga(III) (Figure 2). The most favorable Sc(III) and Y(III) *surface* sites share the same CB carboxylate ligand configuration that had been noted to be unfavorable for Ga(III) (Figure 2 and Supporting Information Table S3). In experiments, partial surface oxidation was observed to facilitate formation of a Y shell on InP QD cores that led to minor red shifting of the absorption spectrum.[91] Our calculated HOMO-LUMO gaps on the most favorable Y-doped InP MSC configurations are consistent with these observations (Supporting Information Figure S10).[91] Thus, tailoring local ligand coordination can be expected to be useful in influencing the sites and favorability of M(III) dopants of different sizes.

**4b. Charge Balancing Effects with M(II) Doping.**

Since addition of an M(II) dopant requires removal of a surface ligand to maintain charge neutrality[30], it can be expected to be even more ligand binding-mode- and coordination-dependent (see Secs. 2-3). For example, removal of a CB ligand is more straightforward since it is localized to a single cation site, and thus its removal does not destabilize neighboring In *surface* (i.e., P = 1 or P = 2) sites. First evaluating energetics for Cd(II) doping, we indeed observe a strong influence of binding mode of the removed ligand on energetics. As expected, doping sites formed by removing CB ligands are on average 12 kcal/mol more favorable than either of the sites formed by removing ABA/ABF or SB ligands (Figure 3 and Supporting Information Table S13). This trend can also be attributed to a reduced energetic penalty for ligand removal, as increased favorability of Cd(II)-substitution is reasonably correlated ($R^2$=0.39) to MSC In(III) sites that were previously observed to have lower rigid ligand dissociation energies[61] (Supporting Information Figure S11). For a more consistent comparison, we also define a rigid Cd(II) doping energy (i.e., without letting the environment around the cation site reorganize when Cd(II) is inserted). Comparing this quantity to the rigid ligand dissociation energy increases the correlation ($R^2$ = 0.69), suggesting that Cd(II) doping is influenced both by the energetic penalty for ligand removal and the Cd(II)-local relaxation of the MSC (Supporting Information Figure S11).

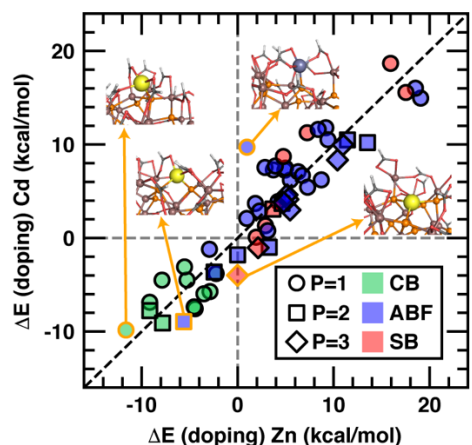

**Figure 3.** ΔE (doping) for Zn(II) vs Cd(II) in the InP MSC colored by the binding mode of the ligand removed: chelating bidentate (CB, in green), asymmetric bridging facing (ABF, in blue), and symmetric bridging (SB, in red). The number of coordinating phosphorus atoms is indicated by the symbol for P = 1 (circles), P = 2 (squares), and P = 3 (diamonds). Inset structures of relaxed, doped-MSCs are shown with dopants shown as enlarged spheres. Atoms are colored as Cd in yellow, Zn in gray, In in brown, O in red, C in gray, and H in white. A black dashed parity line, and gray dashed ΔE (doping) = 0 kcal/mol lines are also shown.

In contrast to the M(III) dopants where we observed a significant distinction between *surface* (i.e., P = 1 or P = 2) and *interior* (i.e., P = 3) site-dependent doping energetics, Cd(II) dopants have comparable average energetics for these two types of sites (Supporting Information Table S8 and Table S14). At least one favorable (i.e., exothermic) Cd(II) site is observed for both the *surface* P = 1 (-9.8 kcal/mol) and *interior* P = 3 (-4.0 kcal/mol) MSC sites (Figure 3 and Supporting Information Table S15). Despite comparable average values, the most favorable sites are indeed again on the surface of the MSC (Figure 3). For the 19 sites with exothermic doping energies, 17 correspond to *surface* P = 1 or P = 2 sites that also tend to be coordinated by CB ligands. We conclude that the MSC surface is thus more favorable for Cd(II) doping, which is in agreement with experimental observations[32] (Figure 3 and Supporting Information Table S15).

To determine whether size effects alter trends with M(II) dopants as they did for the isovalent M(III) case, we next evaluated doping energetics with Zn(II) for the most favorable (i.e., *surface* P = 1 or P = 2) sites. The comparable covalent radii of the two cations (i.e., Zn(II):

1.22 Å vs Cd(II): 1.44 Å) leads to similar doping site favorability on average, with differences around 2.3 kcal/mol (Figure 3 and Supporting Information Table S15). This suggests that local ligand coordination plays a larger role than cation identity in M(II) doping, although we note that this observation can be expected from the comparable size (i.e., covalent radius within 0.2 Å) of the two dopants to each other and to the In(III) (i.e., 1.42 Å) cation (Supporting Information Tables S10–S11). Instead focusing on the sites with the largest differences between Zn(II) and Cd(II), we observe that the greatest difference occurs for a surface site where the smaller Zn(II) cation leads to greater structural rearrangement around the surrounding MSC, suggesting some differences that could result in limiting cases due to size differences (Figure 3 and Supporting Information Figure S12). Analysis of the M(II)-doped MSC HOMO-LUMO gap for both Zn(II) and Cd(II) indicates a slightly reduced gap when doping occurs at the favorable sites (i.e., a slight red-shift of the absorbance spectrum) in agreement with both prior calculations and UV-Vis experiments (Supporting Information Figure S13).[30-31, 35] Overall, M(II) dopant identities are similar enough that the predominant distinguishing effects on energetics appear to be due to differences in the ligand type removed during the doping process.

**4c. Overall Trends with Single and Increasing Dopant Concentrations.**

To provide insight into whether relatively favorable and unfavorable sites are concentrated on the surface and, if so, whether such sites are universal across dopants, we take a global view of the M(II) and M(III) doping energetics on the MSC surface. While for M(II) doping, both (i.e., Zn and Cd) dopants are readily incorporated over most surface sites, the most favorable sites are more localized to the top of the pseudo C2 symmetry axis of the MSC for Zn(II) than Cd(II) (Figure 4). Cd(II) conversely favorably dopes over more surface sites than Zn(II), which rationalizes experimental observations of facile Cd(II) doping (Figure 4).[30, 35]

Comparing M(II) dopants to Ga(III) reveals that much less of the MSC surface is favorable for Ga(III) doping (Figure 4). Indeed, the favorable surface sites for Ga(III) doping remain buried underneath thermoneutral and unfavorable *surface* sites (Figure 4). This observation is supported by experiments where the shorter-chain Ga(II) acetate was necessary to exchange dopants into MA-protected MSC.[35] Unlike Ga(III), Y(III) doping energetics are mostly thermoneutral across the surface along with one or two slightly favorable sites. Thus, although doping with Y(III) is feasible, Y(III) is likely to be localized to specific sites on the MSC more than M(II) dopants (Figure 4). Accounting for both single-doping site energetics by elements along with the accessibility of the most favorable sites thus explains why some dopants (e.g., Ga(III)) have been more challenging to incorporate in both the MSC and larger QDs.

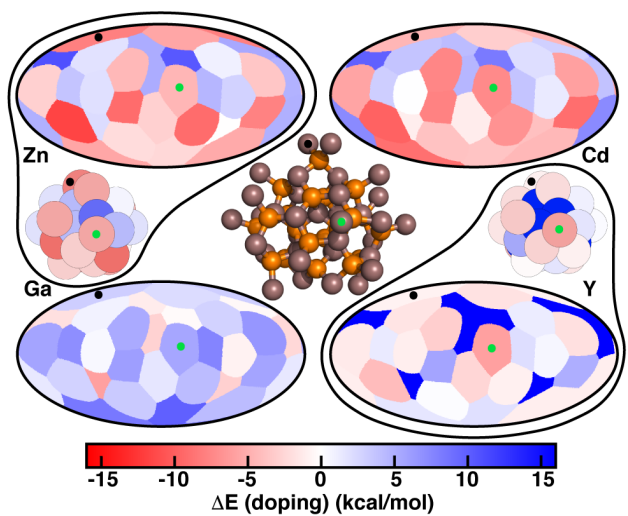

**Figure 4.** Mollweide (equal area) map projections of the surface energetics of Zn(II), Cd(II), Ga(III), and Y(III) dopants in the MSC colored according to the bottom colorbar (in kcal/mol). The scale of the ΔE (doping) is limited to 16 kcal/mol and centered at zero, capping a subset of interior Y(III) doping energetics that exceed this range. For consistent comparisons between M(II) and M(III) dopants, only the most favorable ΔE (doping) value was reported for cases where multiple options of ligand removal were possible (i.e., for M(II)). Green and black points indicate 2 identical sites across the MSC for comparison.

Given the differences apparent for single dopant sites, we next evaluated the energetics of increasing dopant concentrations in the MSC. This also allows us to make comparisons to experiments that frequently study dopant concentrations above one per cluster.[30, 35-36] Starting from M(II) dopants, we predict that additional Cd(II) dopants are favorable at the MSC surface (Figure 5 and Supporting Information Table S16). The additional most favorable sites at higher concentrations largely coincide with the energies that would have been predicted by independently summing the lowest-energy single dopant sites (Figure 5 and Supporting Information Table S16). Nevertheless, some exceptions are noteworthy, with a deviation from energetics predicted by the sum of single-site dopant energies first occurring with five Cd(II) dopants (i.e, 14 mol% Cd). This deviation coincides with the first removal of a bridging ligand instead of a CB ligand (Supporting Information Table S16 and Figure S14). For the more extensively Cd(II)-doped MSC, there is evidently less flexibility to rearrange and stabilize the surface around this missing ligand relative to the singly-doped case (Supporting Information Table S16 and Figure S14).

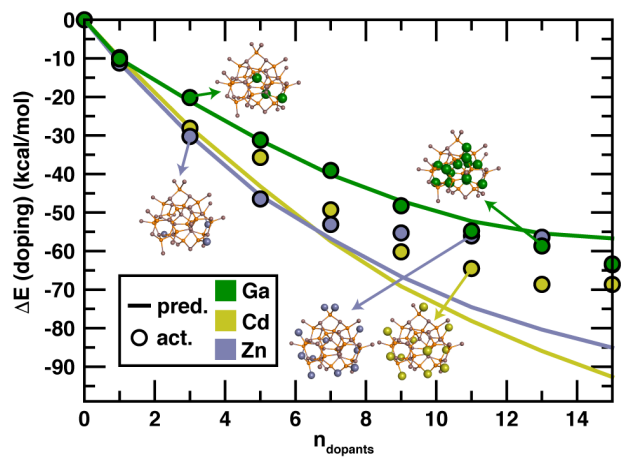

**Figure 5.** Cumulative ΔE (doping) in kcal/mol for lowest-energy sites sampled versus number of dopants, $n_{dopants}$. The predicted doping energetics (pred., shown as solid lines) are based on the sum of the equivalent single site ΔE (doping). The $n_{dopants}$ is the number of dopant atoms in the MSC out of a total of 37 In sites both in the core and on the surface. Insets show both low and high concentration-configurations for both M(II) and Ga(III).

Experimentally, a maximum of 16 Cd(II) dopants were added to the MSC before it transformed phase, likely into the tetragonal crystal structure of $Cd_3P_2$.[30] Notably, the calculated energetic favorability of doping rapidly diminishes from 11 to 15 Cd(II) dopants (i.e., 30 to 41 mol% Cd) in the MSC, and thus our calculations indicate that further doping would require more significant rearrangement like that observed experimentally (Figure 5 and Supporting Information Figure S15). With increasing Cd(II) doping, strongly-bound bridging ligand modes increase in concentration at the MSC surface and saturate at 15 Cd(II), indicating why reconstruction of the MSC is required for additional Cd(II) doping (Supporting Information Figure S14). This increased Cd(II) doping also decreases HOMO-LUMO gaps, indicating the source of red-shifts observed experimentally in low-energy UV-Vis absorption spectra[30] (Figure 6 and Supporting Information Figure S16). Thus, our calculations suggest Cd(II) doping at low- to moderate- concentrations have minimal cooperative effects between single doping sites, and favorability is only limited at the highest surface concentrations.

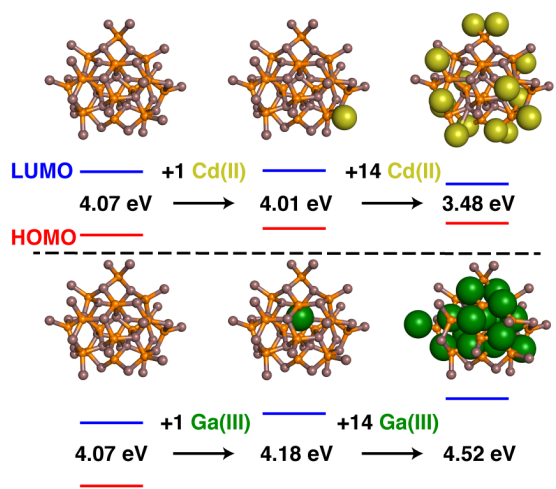

**Figure 6.** Illustration of HOMO-LUMO gap and structural shifts in the MSC when Cd(II) (upper) and Ga(III) (lower) are doped to higher concentrations.

Since we observed limited differences of single dopant energetics with Zn(II) and Cd(II), we would expect to observe a similar concentration dependence for Zn(II). The Zn(II) multiple-dopant energetics at low- to moderate- concentrations are indeed additive in a similar fashion to Cd(II), with at least some marginal benefit of ΔE (doping) observed up to 15 dopants (Figure 5). However, in the moderate to high-dopant concentration regime (i.e., seven to 13 dopants), the marginal benefit of adding Zn(II) is significantly smaller than Cd(II) (Figure 5 and Supporting Information Figure S15). While the number of dopants at which Zn(II) starts to deviate from the sum of single-site dopant energies, it in fact again coincides with the removal of the first bridging ligand at seven substitutions, which is at a higher dopant concentration from the five dopants for Cd(II) (Supporting Information Table S16). This is consistent with experimental observations of Zn-doped MSC phase transformation and rearrangement at a smaller number (i.e., between five and nine) of Zn(II) per MSC than Cd(II).[35] In comparison to Cd(II), our calculations indicate Zn(II) also has a smaller effect on the HOMO-LUMO gap of the MSC, consistent with experimental spectral shifts[35] (Supporting Information Figure S16). Therefore, small differences between single dopant energetics of Cd(II) and Zn(II) become more pronounced at higher concentrations, with Zn(II) showing more evidence of cooperativity between doping sites.

Since M(II) multiple-site doping was favorable with limited to moderate cooperativity only observed at high concentrations as a change in the type of ligand removed changed, we expected M(III) cations with favorable single-site dopant energetics (i.e., Ga(III)) to exhibit favorable doping to higher concentrations. Indeed, addition of multiple Ga(III) to MSCs is consistently favorable to high concentrations and follows trends expected from single-site dopants more closely (Figure 5 and Supporting Information Figure S15). In fact, the multiple-dopant energetics for Ga(III) at high concentrations are actually slightly more favorable than would be predicted alone by the sum of the independent single-site energetics (Figure 5). Such observations suggest the possibility of favorable cooperative effects in the moderate-doping regime of seven or more dopants (Figure 5). This computational observation of thermodynamic favorability contrasts with experimental findings where only three Ga(III) were successfully doped in the MSC with acetate Ga(III) precursors.[35] This difference is likely caused by increased kinetic barriers[35] for diffusion of Ga(III) into the MSC core where it is most favorable (Figure 2). Nevertheless, the predicted favorable energetics for higher Ga(III) concentrations suggest possible further incorporation of Ga(III) into the MSC to form a core-shell Ga-In MSC structure, which calculations indicate will also have an increased HOMO-LUMO gap relative to the InP MSC (Figure 6 and Supporting Information Figure S16 and Table S16).

Over all three cases, analyzing the summed single site ΔE (doping) and multiple doped ΔE (doping) configurations favored reveals that the sequentially added most favorable sites in the multiply-doped structures largely remain consistent with the single-dopant analysis. Where exceptions are observed, they are smallest in Ga(III) and increase slightly with Cd(II) and Zn(II). These distinctions in the M(II) dopants are expected due to decreasing numbers of CB ligands on the MSC from charge-balancing ligand loss with M(II) (Supporting Information Figures S14 and S17). Thus, rapid screening of single sites will provide critical insight into understanding dopant favorability both globally and with respect to local structural variation, especially in isovalent doping.

## 5. Conclusions

Introducing dopants to quantum dots as a means of tuning photoluminescent properties has attracted intense experimental and computational interest. Nevertheless, it is difficult to know *a priori* the means of rationally tuning properties of the quantum dot or where dopants will reside with increasing concentration. Furthermore, even after introduction of dopants is established, characterization of the new nanostructured material can be challenging. We addressed these limitations by carrying out extensive large-scale electronic structure calculations of the elementary thermodynamics of quantum dot doping on a model magic size cluster of InP that had been structurally characterized. Through these large-scale electronic structure simulations, we identified the thermodynamic favorability of dopants to be both specific to the sites within the MSC and to the nature of the dopant atom. These observations motivated us to develop maps of the most and least favorable doping sites. For isovalent doping (i.e., Y(III)/Sc(III) or Ga(III), we observed stronger sensitivity of the predicted energetics to the type of ligand orientation on the surface than to the dopant type, but interior doping was favorable only in the case of smaller Ga(III). Limited experimental observations of interior doping of Ga(III) suggests the process would be limited by kinetic barriers to diffusion. For the M(II) (i.e., Zn or Cd) dopants, doping was always favorable at the surface, but the type of ligand removed during the doping reaction was most critical.

We also observed that limited cooperativity occurred with dopants up to moderate concentrations, suggesting that rapid single-dopant estimations of favorability could be carried out as an efficient means of determining the effect of multiple dopants. Predictions of when marginal benefit of doping energetically was diminished coincided with experimental observations of phase transformations (e.g., in Zn(II) or Cd(II)). While some of our predictions were surprising (e.g., of favorable interior doping of Ga(III)), we rationalized our observations by corroborating the blue or red shift of frontier orbital energetics with doping in comparison to available experimental results. Overall, this work highlights the importance of ligand chemistry and surface heterogeneity in determining paths to favorable doping in quantum dots. We expect the strategy demonstrated here to be useful in other III-V and II-VI quantum dot systems, especially those with similar (i.e., carboxylate) ligand chemistries.

## ASSOCIATED CONTENT

**Supporting Information**. MSC In sites grouped by number of In-P bonds; Summary of terminology for ligand binding modes ; Enumerated In sites in the MSC; Comparison across doping sites with Cd on the MSC; Comparison of formate vs. acetate ligands on the MSC; Energetics of Zn doping with formate vs. acetate ligands on the MSC; Comparison of functionals on doping energies on the MSC; Basis set comparison on doping energies on the ESC; Enumerated In sites in the ESC; Basis set comparison energies in the ESC; Comparison of solvent corrections on the MSC; Solvent corrections energies on the MSC; Stacked histograms of M(III) doping energetics; Energetics of bidentate swapping of ligands; MSC single M(III) energetics by number of M-P bonds; Empirical Lewis acidities of metal cations; MSC single M(III) doping energetics; Covalent radii of elements; M(III)-P bond distances at each M-P coordination number; In-P bond distances at each In-P coordination number; HOMO-LUMO gaps of single M(III) doped MSCs; M-P bond distance statistics at each In-P coordination number; MSC single M(II) statistics by ligand binding mode; Cd(II) doping energetics versus rigid ligand dissociation; MSC single M(II) statistics by number of M-P bonds; MSC single M(II) doping energetics; RMSD and energetic comparison across M(II) dopants; HOMO-LUMO gaps of single M(II) doped MSCs; Energetics of multiple M(II) and M(III)-doped MSCs; Geometries of multiple M(II) and M(III)-doped MSCs energetics; Marginal addition energetics of multiple M(II) and M(III)-doped MSCs; HOMO-LUMO gaps of multiple M(II) and M(III)-doped MSCs; Single vs multiple M(II) and M(III)-doped MSCs energetics. (PDF)

Initial and final structures of doped MSC and precursor molecules. (ZIP)

This material is available free of charge.

## AUTHOR INFORMATION

**Corresponding Author**

*email: hjkulik@mit.edu phone: 617-253-4584

**Notes**

The authors declare no competing financial interest.

## ACKNOWLEDGMENT

**Corresponding Author**

*email: hjkulik@mit.edu phone: 617-253-4584

**Notes**

The authors declare no competing financial interest.

This work was primarily supported by the U.S. Department of Energy DE-SC0012702. The software tool development was also supported by the Office of Naval Research under grant number N000014-20-1-2150. Early stages of this work were supported by the National Science Foundation under grant number ECCS-1449291. This work made use of Department of Defense HPCMP computing resources. This work was also carried out in part using computational resources from the Extreme Science and Engineering Discovery Environment (XSEDE), which is supported by National Science Foundation grant number ACI-1548562. H.J.K. holds a Career Award at the Scientific Interface from the Burroughs Wellcome Fund, an AAAS Marion Milligan Mason Award, and a Alfred P. Sloan Fellowship in Chemistry, which supported this work. The authors thank Daniel R. Harper, Aditya Nandy, and Adam H. Steeves for providing a critical reading of the manuscript.

**For Table of Contents Use Only**

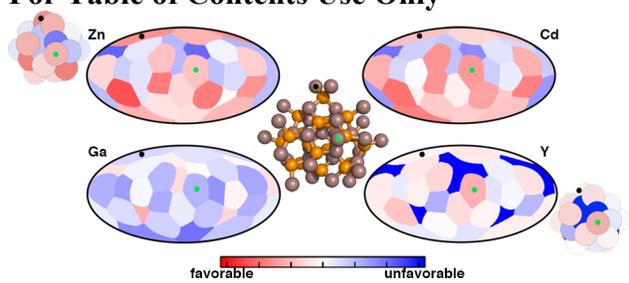

# Supporting Information for
## *Mapping the Electronic Structure Origins of Surface- and Chemistry-Dependent Doping Trends in III-V Quantum Dots*

Michael G. Taylor[1] and Heather J. Kulik[1]

[1]Department of Chemical Engineering, Massachusetts Institute of Technology, Cambridge, MA 02139

**Contents**

**Table S1.** MSC In sites grouped by number of In-P bonds (P) as well as by type of In site: surface (1 or 2 In-P bonds, only), interior (3 In-P bonds only), and bulk (4 In-P bonds only). The total count for each type/number of In sites is also reported.

|  | P | In indices | total |
|---|---|---|---|
| Surface | 1 | 1,2,3,5,6,12,20,23,24,25,26,27,34,36,36 | 16 |
|  | 2 | 4,8,11,17,29,32 | 6 |
| Interior | 3 | 7,13,14,15,18,30,31,33 | 8 |
| Bulk | 4 | 9,10,16,19,21,22,28 | 7 |

**Table S2.** Summary of terminology for ligand binding modes used in literature.

| this work | refs[1-2] | ref[3] | ref[4] | ref[5] | ref[6] |
|---|---|---|---|---|---|
| chelating bidentate | chelating bidentate | chelate | chelating bidentate | bidentate | bidentate |
| symmetric bridging | bridging bidentate | tilted | bridging bidentate | symmetric bridging | symmetric bridging |
| asymmetric bridging | chelating bridging bidentate | bridging | chelating bridging bidentate | asymmetric bridging | asymmetric bridging / asymmetric bridging + dative |

**Table S3.** MSC In coordination numbers by In site index: In-P bonds (P), chelating bidentate carboxylates (CB), symmetric bridging (SB) ligands, asymmetric bridging facing (ABF) ligands, and asymmetric bridging away (ABA) ligands. The min, max, average, and standard deviation values are shown at the bottom of the table. The bond valence (BV), which is the sum of the Mayer bond order is also given. Multiwfn[7] was used to compute Mayer bond orders.[8-9]

| In index | P | CB | SB | ABF | ABA | BV |
|---|---|---|---|---|---|---|
| 1 | 1 | 0 | 0 | 2 | 2 | 2.43 |
| 2 | 1 | 1 | 0 | 1 | 2 | 2.54 |
| 3 | 1 | 1 | 0 | 1 | 2 | 2.49 |
| 4 | 2 | 1 | 0 | 0 | 2 | 2.49 |
| 5 | 1 | 1 | 0 | 1 | 2 | 2.55 |
| 6 | 1 | 1 | 0 | 1 | 2 | 2.54 |
| 7 | 3 | 0 | 0 | 1 | 0 | 2.50 |
| 8 | 2 | 0 | 0 | 2 | 1 | 2.46 |
| 9 | 4 | 0 | 0 | 0 | 0 | 2.29 |
| 10 | 4 | 0 | 0 | 0 | 0 | 2.22 |
| 11 | 2 | 0 | 1 | 1 | 1 | 2.47 |
| 12 | 1 | 0 | 1 | 2 | 1 | 2.48 |
| 13 | 3 | 0 | 0 | 1 | 0 | 2.44 |
| 14 | 3 | 0 | 1 | 0 | 0 | 2.53 |
| 15 | 3 | 0 | 0 | 1 | 0 | 2.52 |
| 16 | 4 | 0 | 0 | 0 | 0 | 2.28 |
| 17 | 2 | 0 | 0 | 2 | 1 | 2.45 |
| 18 | 3 | 0 | 0 | 1 | 0 | 2.41 |
| 19 | 4 | 0 | 0 | 0 | 0 | 2.30 |
| 20 | 1 | 1 | 0 | 1 | 2 | 2.51 |
| 21 | 4 | 0 | 0 | 0 | 0 | 2.25 |
| 22 | 4 | 0 | 0 | 0 | 0 | 2.29 |
| 23 | 1 | 1 | 1 | 0 | 2 | 2.51 |
| 24 | 1 | 0 | 1 | 2 | 1 | 2.48 |
| 25 | 1 | 0 | 0 | 3 | 1 | 2.46 |
| 26 | 1 | 0 | 0 | 3 | 1 | 2.47 |
| 27 | 1 | 1 | 1 | 1 | 1 | 2.54 |
| 28 | 4 | 0 | 0 | 0 | 0 | 2.27 |
| 29 | 2 | 0 | 1 | 1 | 1 | 2.50 |
| 30 | 3 | 0 | 0 | 1 | 0 | 2.49 |
| 31 | 3 | 0 | 0 | 1 | 0 | 2.51 |
| 32 | 2 | 1 | 0 | 0 | 2 | 2.47 |
| 33 | 3 | 0 | 0 | 1 | 0 | 2.51 |
| 34 | 1 | 1 | 1 | 0 | 2 | 2.54 |
| 35 | 1 | 0 | 0 | 2 | 2 | 2.44 |
| 36 | 1 | 1 | 0 | 1 | 2 | 2.53 |
| 37 | 1 | 1 | 0 | 1 | 2 | 2.51 |
| MIN | 1 | 0 | 0 | 0 | 0 | 2.22 |
| MAX | 4 | 1 | 1 | 3 | 2 | 2.55 |
| AVERAGE | 2.16 | 0.32 | 0.22 | 0.95 | 0.95 | 2.45 |
| STD DEV | 1.19 | 0.47 | 0.42 | 0.85 | 0.88 | 0.09 |

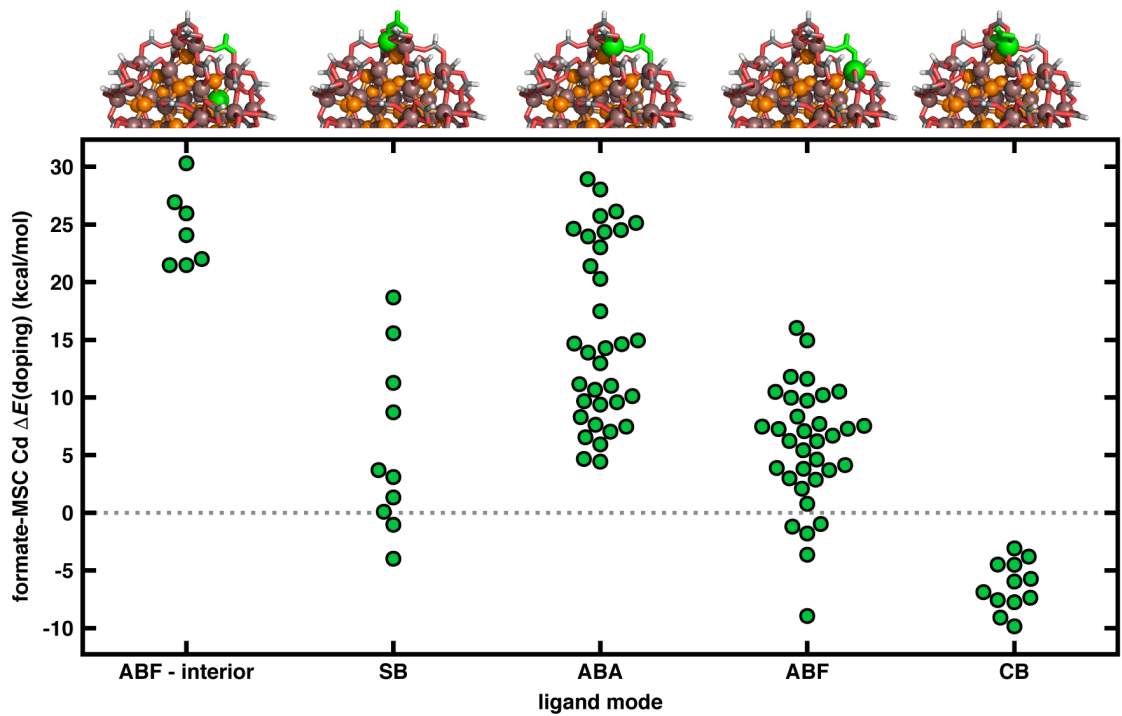

**Figure S1.** (upper) Images display five ligand binding modes. From left to right the ligand modes are: asymmetric bridging facing interior doping (ABF – interior), symmetric bridging (SB), asymmetric bridging away (ABA), asymmetric bridging face (ABF), and chelating bidentate (CB). Green spheres indicate the In that is replaced by Cd, while green sticks highlight the ligand deleted before re-optimizing. (lower) Comparison of Cd $\Delta E$(doping) (in kcal/mol) by five binding modes of the ligands. We note that the ABF – interior and ABA binding modes show $\Delta E$(doping) values only above zero, and are therefore largely unfavorable. The SB, ABF, and CB binding modes show at least some values less than zero indicating favorable Cd doping.

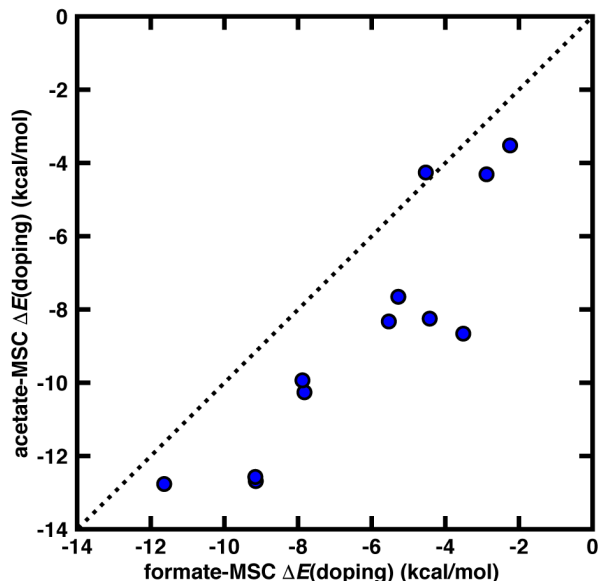

**Figure S2.** Comparison of ΔE(doping) of Zn at chelating bidentate sites (in kcal/mol) between the acetate-MSC with respect to the formate-MSC. We note with acetate vs. formate, the overall trends in site-specific doping favorability was generally maintained, along with favorability of doping (-8.6 kcal/mol with formate vs. -6.2 kcal/mol with acetate).

**Table S4.** ΔE(doping) of Zn at chelating bidentate sites (in kcal/mol) between the acetate-MSC with respect to the formate-MSC. We note acetate energetics are more favorable on average by 2.4 (kcal/mol) for Zn.

| In index doping | formate-MSC ΔE(doping) (kcal/mol) | acetate-MSC ΔE(doping) (kcal/mol) | ΔΔE(doping) (formate-acetate) (kcal/mol) |
|---|---|---|---|
| 2 | -2.9 | -4.3 | 1.4 |
| 3 | -5.3 | -7.7 | 2.4 |
| 4 | -7.8 | -10.3 | 2.4 |
| 5 | -11.6 | -12.8 | 1.1 |
| 6 | -3.5 | -8.7 | 5.1 |
| 20 | -2.2 | -3.5 | 1.3 |
| 23 | -7.9 | -9.9 | 2.1 |
| 27 | -9.2 | -12.6 | 3.4 |
| 32 | -9.1 | -12.7 | 3.5 |
| 34 | -5.5 | -8.3 | 2.8 |
| 36 | -4.4 | -8.2 | 3.8 |
| 37 | -4.5 | -4.3 | -0.3 |
| MIN | -11.6 | -12.8 | -0.3 |
| MAX | -2.2 | -3.5 | 5.1 |
| AVERAGE | -6.2 | -8.6 | 2.4 |

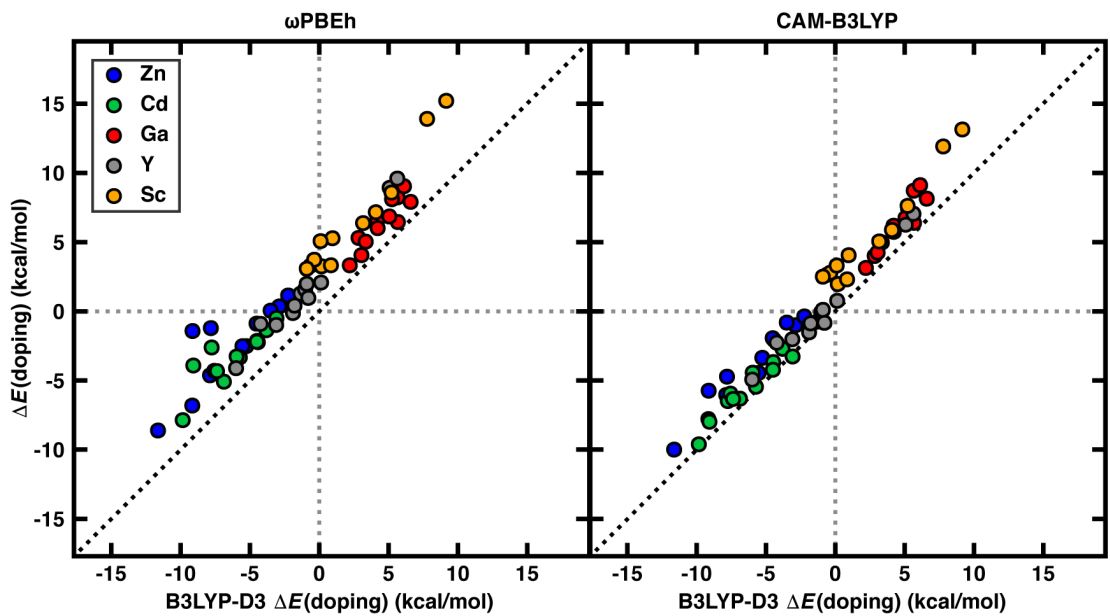

**Figure S3.** Comparison of Δ*E*(doping) for all dopants (in kcal/mol) between the B3LYP-D3 functional with ωPBEh (left) and CAM-B3LYP (right) functionals over identical In-doping positions from chelating bidentate-bound locations in the formate-In37 MSC according to energetics evaluated on B3LYP-D3 relaxed structures. Symbols are colored by dopant identity following inset legend. Somewhat larger deviations (i.e., a positive shift) from parity are observed for ωPBEh than for CAM-B3LYP, although all functionals have consistent relative dopant favorability and relative site preference across dopants.

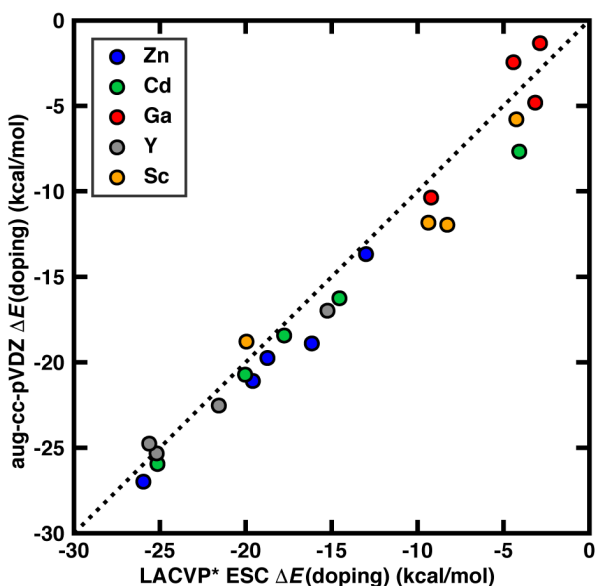

**Figure S4.** Basis set comparison for Δ*E*(doping) (in kcal/mol) between the aug-cc-pVDZ and LACVP* basis set evaluated on the ESC across all dopants tested at chelating-bidentate sites on B3LYP-D3/LACVP*-relaxed structures. We note that both relative site preference among each dopant is maintained, along with trends between average dopant favorability.

**Table S5.** Enumeration of doping configurations in an early stage cluster (ESC). The initial early stage cluster (ESC) with formate ligands (In$_6$P(O$_2$CH)$_{15}$) was generated by replacing the acetate ligands from prior AIMD.[4] We note each ligand with a bridging binding mode; asymmetric bridging face (ABF), asymmetric bridging away (ABA), and symmetric bridging (SB) have 2 ligand doping configurations. Chelating bidentate ligands (CB) only have one ligand doping configuration.

| doping configuration number | ligand binding mode | In index doping | ligand deleted |
|---|---|---|---|
| 1 | ABF | 3 | 10 |
| 2 | SB | 3 | 12 |
| 3 | ABF | 3 | 8 |
| 4 | CB | 4 | 4 |
| 5 | ABA | 4 | 6 |
| 6 | CB | 4 | 7 |
| 7 | ABA | 4 | 8 |
| 8 | ABA | 4 | 9 |
| 9 | SB | 5 | 12 |
| 10 | SB | 5 | 13 |
| 11 | CB | 6 | 15 |
| 12 | ABA | 1 | 10 |
| 13 | CB | 1 | 11 |
| 14 | SB | 1 | 1 |
| 15 | ABF | 1 | 5 |
| 16 | ABF | 1 | 9 |
| 17 | SB | 2 | 13 |
| 18 | SB | 2 | 14 |
| 19 | SB | 2 | 1 |
| 20 | ABA | 2 | 5 |
| 21 | SB | 5 | 2 |
| 22 | CB | 5 | 3 |
| 23 | SB | 6 | 14 |
| 24 | SB | 6 | 2 |
| 25 | ABF | 6 | 6 |

**Table S6.** aug-cc-pVDZ vs. LACVP* basis set ESC Δ$E$(doping) (in kcal/mol) across all dopants tested at chelating-bidentate sites on B3LYP-D3/LACVP*-relaxed structures. We note very close agreement across basis sets for most energetically-favored doping locations for each dopant. We make basis set comparisons on an ESC that contains one P and six surrounding In. For the ESC, there are five CB ligands and 10 SB or AB ligands, producing 25 unique possibilities for removing formate ligands from the surface during M(II) doping, all of which we evaluate.

| doping metal | In index doping | ligand deleted | aug-cc-pVDZ Δ$E$(doping) (kcal/mol) | LACVP* Δ$E$(doping) (kcal/mol) | ΔΔ$E$(doping) (aug-cc-pVDZ - LACVP*) (kcal/mol) |
|---|---|---|---|---|---|
| Cd | 1 | 11 | -16.3 | -14.5 | -1.7 |
| Cd | 4 | 4 | -20.7 | -20.0 | -0.7 |
| Cd | 4 | 7 | -7.7 | -4.1 | -3.6 |
| Cd | 5 | 3 | -18.4 | -17.8 | -0.7 |
| Cd | 6 | 15 | -25.9 | -25.1 | -0.8 |
| Zn | 1 | 11 | -21.1 | -19.6 | -1.5 |
| Zn | 4 | 4 | -13.7 | -13.0 | -0.7 |
| Zn | 4 | 7 | -18.9 | -16.2 | -2.7 |
| Zn | 5 | 3 | -19.8 | -18.7 | -1.0 |
| Zn | 6 | 15 | -27.0 | -26.0 | -1.0 |
| Ga | 1 | - | -2.5 | -4.4 | 2.0 |
| Ga | 4 | - | -1.3 | -2.9 | 1.5 |
| Ga | 5 | - | -4.8 | -3.2 | -1.6 |
| Ga | 6 | - | -10.4 | -9.2 | -1.1 |
| Y | 1 | - | -25.3 | -25.2 | -0.1 |
| Y | 4 | - | -17.0 | -15.2 | -1.7 |
| Y | 5 | - | -24.8 | -25.6 | 0.9 |
| Y | 6 | - | -22.5 | -21.6 | -1.0 |
| Sc | 1 | - | -12.0 | -8.3 | -3.7 |
| Sc | 4 | - | -11.8 | -9.4 | -2.5 |
| Sc | 5 | - | -18.8 | -20.0 | 1.2 |
| Sc | 6 | - | -5.8 | -4.3 | -1.5 |
| | | MIN | -27.0 | -26.0 | -3.7 |
| | | MAX | -1.3 | -2.9 | 2.0 |
| | | AVERAGE | -15.7 | -14.7 | -1.0 |

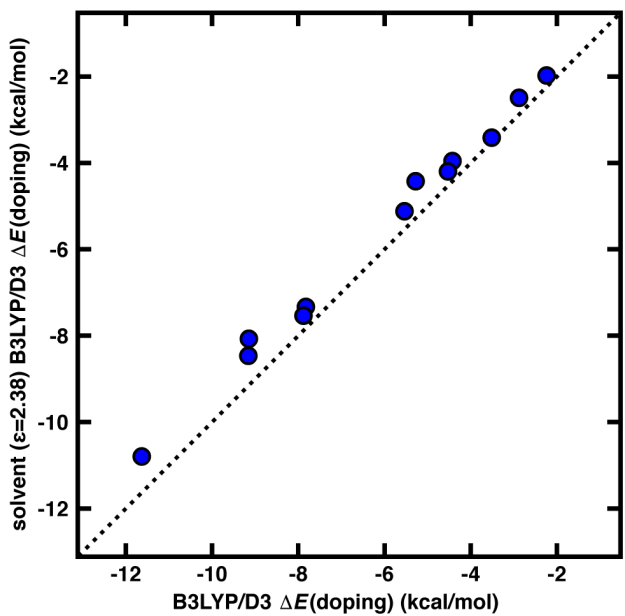

**Figure S5.** Comparison of Δ*E*(doping) (in kcal/mol) of implicit solvent-corrected doping of Zn at chelating bidentate sites (in kcal/mol) with respect to the uncorrected B3LYP-D3 in uncorrected B3LYP-D3 relaxed formate-In37 MSCs. We note only minor deviation in the trend in Zn doping site preference accounting for solvent interactions.

**Table S7.** Δ*E*(doping) (in kcal/mol) of implicit solvent-corrected doping of Zn at chelating bidentate sites (in kcal/mol) and uncorrected B3LYP-D3 in uncorrected B3LYP-D3 relaxed formate-In37 MSCs. We note agreement within 1.1 (kcal/mol) and slightly less favorable doping energetics (0.5 (kcal/mol)) with solvent corrections.

| In index doping | B3LYP/D3 ΔE(doping) (kcal/mol) | solvent (ε=2.38) B3LYP/D3 ΔE(doping) (kcal/mol) | ΔΔE(doping) (B3LYP/D3 - solvent) (kcal/mol) |
|---|---|---|---|
| 2 | -2.9 | -2.5 | -0.4 |
| 3 | -5.3 | -4.4 | -0.9 |
| 4 | -7.8 | -7.3 | -0.5 |
| 5 | -11.6 | -10.8 | -0.8 |
| 6 | -3.5 | -3.4 | -0.1 |
| 20 | -2.2 | -2.0 | -0.3 |
| 23 | -7.9 | -7.5 | -0.3 |
| 27 | -9.2 | -8.5 | -0.7 |
| 32 | -9.1 | -8.1 | -1.1 |
| 34 | -5.5 | -5.1 | -0.4 |
| 36 | -4.4 | -4.0 | -0.5 |
| 37 | -4.5 | -4.2 | -0.3 |
| MIN | -11.6 | -10.8 | -1.1 |
| MAX | -2.2 | -2.0 | -0.1 |
| AVERAGE | -6.2 | -5.6 | -0.5 |

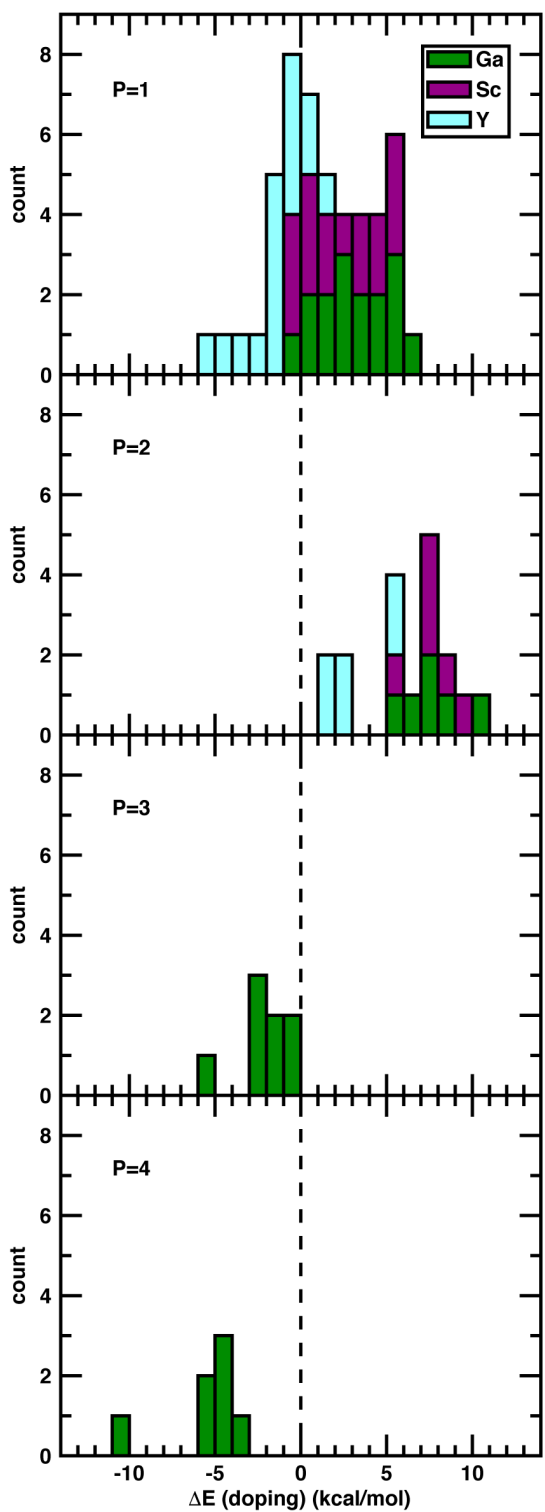

**Figure S6.** Stacked histograms (bin width: 1 kcal/mol) of ΔE (doping) for M(III) dopants (Sc in purple, Ga in dark green, and Y in cyan) into the InP MSC grouped by increasing P-coordination (top to bottom). The dashed vertical line at zero indicates thermoneutral ΔE (doping).

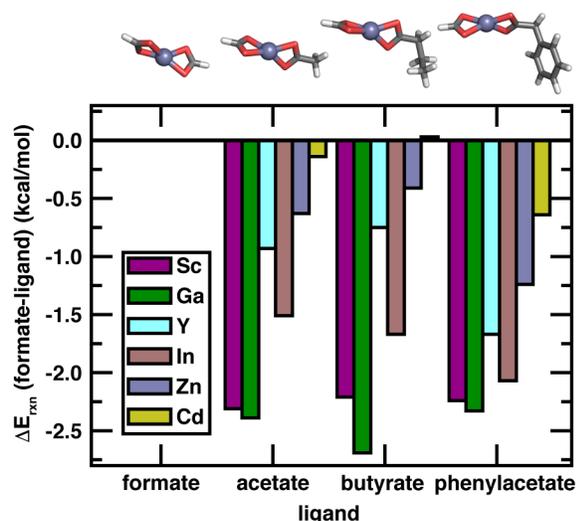

**Figure S7.** $\Delta E_{rxn}$ (kcal/mol) for bidentate swapping of ligands on metal ion precursors. $\Delta E_{rxn}$ is the electronic energy over the reaction $HO_2CR + M(O_2CH)_x \rightarrow HO_2CH + M(O_2CH)_{x-1}(O_2CR)_1$ where R=$CH_3$,$CH_2CH_2CH_3$, and $CH_2Ph$, M represents any of the metals (Sc, Ga, Y, In, Zn, or Cd), and $x$ is the oxidation state of the metal (e.g., for Ga(III) $x$=3). Example selected Zn precursor structures are shown at top with Zn in light purple, O in red, C in gray, and H in white. Note that the largest magnitude $\Delta E_{rxn}$ from formate comes from Ga(III) bound with butyrate and is -2.6 kcal/mol.

**Table S8.** MSC single M(III) (Ga, Sc, and Y) min, max, average, and standard deviation of $\Delta E$ doping in kcal/mol for each P-coordination number. In indices for the max and min $\Delta E$ (doping) are added for each defined P-coordination and metal.

| metal | P | MIN In index | MAX In index | MIN | MAX | AVERAGE | STD DEV |
|---|---|---|---|---|---|---|---|
| Ga | 1 | 12 | 6 | -0.8 | 6.6 | 3.0 | 2.1 |
| Sc | 1 | 37 | 12 | -0.9 | 5.9 | 2.3 | 2.3 |
| Y | 1 | 37 | 25 | -6.0 | 1.2 | -1.6 | 1.8 |
| Ga | 2 | 4 | 17 | 5.7 | 10.0 | 7.5 | 1.6 |
| Sc | 2 | 29 | 32 | 5.6 | 9.2 | 7.6 | 1.2 |
| Y | 2 | 11 | 32 | 1.3 | 5.6 | 2.9 | 1.9 |
| Ga | 3 | 15 | 31 | -5.9 | -0.4 | -2.3 | 1.8 |
| Sc | 3 | 18 | 7 | 29.9 | 36.0 | 32.3 | 1.8 |
| Y | 3 | 18 | 14 | 23.8 | 33.9 | 29.7 | 4.1 |
| Ga | 4 | 19 | 9 | -10.1 | -3.9 | -5.4 | 2.1 |
| Sc | 4 | 28 | 21 | 34.0 | 43.5 | 40.3 | 3.2 |
| Y | 4 | 28 | 19 | 34.0 | 54.1 | 40.5 | 6.3 |

**Table S9.** MSC single M(III) (Ga, Sc, and Y) doping energetics including ΔE doping in kcal/mol (ΔE), HOMO-LUMO gap in eV (gap), and HOMO energy level in eV (HOMO) with reference In indices that are replaced with M(II). The min, max, average, and standard deviation values are shown at the bottom of the table. The HOMO-LUMO gap of the undoped MSC is 4.07 eV, while the HOMO energy level is -5.70 eV.

| In index | Ga ΔE | Ga gap | Ga HOMO | Sc ΔE | Sc gap | Sc HOMO | Y ΔE | Y gap | Y HOMO |
|---|---|---|---|---|---|---|---|---|---|
| 1 | 1.4 | 4.06 | -5.68 | 2.7 | 3.71 | -5.64 | -1.0 | 4.05 | -5.67 |
| 2 | 5.7 | 4.06 | -5.68 | 3.2 | 3.96 | -5.65 | -1.3 | 4.05 | -5.68 |
| 3 | 4.2 | 4.06 | -5.69 | 1.0 | 4.01 | -5.65 | -1.0 | 4.05 | -5.68 |
| 4 | 5.7 | 4.07 | -5.68 | 7.8 | 3.82 | -5.67 | 5.1 | 4.06 | -5.69 |
| 5 | 2.8 | 4.08 | -5.70 | -0.6 | 3.97 | -5.64 | -4.2 | 4.04 | -5.65 |
| 6 | 6.6 | 4.07 | -5.67 | 4.1 | 3.88 | -5.66 | -1.9 | 4.03 | -5.68 |
| 7 | -2.5 | 4.08 | -5.68 | 36.0 | 3.67 | -5.70 | 33.9 | 3.99 | -5.68 |
| 8 | 8.4 | 4.02 | -5.66 | 7.0 | 3.68 | -5.68 | 1.5 | 4.11 | -5.68 |
| 9 | -3.9 | 4.10 | -5.71 | 41.0 | 4.03 | -5.81 | 38.5 | 4.08 | -5.70 |
| 10 | -5.0 | 4.10 | -5.70 | 43.5 | 3.96 | -5.75 | 38.7 | 3.95 | -5.61 |
| 11 | 7.1 | 4.07 | -5.70 | 8.3 | 3.48 | -5.63 | 1.3 | 3.98 | -5.66 |
| 12 | -0.8 | 4.10 | -5.70 | 5.9 | 3.78 | -5.65 | 0.6 | 3.95 | -5.59 |
| 13 | -2.0 | 4.06 | -5.68 | 31.6 | 3.57 | -5.69 | 29.6 | 3.98 | -5.69 |
| 14 | -2.9 | 4.05 | -5.67 | 32.7 | 3.68 | -5.68 | 33.9 | 4.05 | -5.71 |
| 15 | -5.9 | 4.04 | -5.66 | 32.9 | 3.78 | -5.68 | 33.5 | 4.01 | -5.66 |
| 16 | -4.6 | 4.14 | -5.74 | 39.4 | 3.93 | -5.80 | 39.3 | 4.14 | -5.70 |
| 17 | **10.0** | **4.03** | **-5.67** | 7.7 | 3.71 | -5.66 | 2.0 | 4.10 | -5.68 |
| 18 | -0.5 | 4.06 | -5.67 | 29.9 | 3.50 | -5.70 | 23.8 | 3.92 | -5.66 |
| 19 | **-10.1** | **4.18** | **-5.75** | 40.7 | 4.13 | -5.84 | **54.1** | **3.99** | **-5.73** |
| 20 | 3.4 | 4.06 | -5.67 | 5.2 | 4.04 | -5.68 | 0.1 | 4.05 | -5.69 |
| 21 | -5.1 | 4.10 | -5.70 | **43.5** | **3.94** | **-5.73** | 38.9 | 3.88 | -5.49 |
| 22 | -5.1 | 4.12 | -5.73 | 40.1 | 3.95 | -5.79 | 40.0 | 4.09 | -5.62 |
| 23 | 3.1 | 4.08 | -5.68 | 0.2 | 3.75 | -5.68 | -0.8 | 4.05 | -5.68 |
| 24 | 0.4 | 4.10 | -5.70 | 3.7 | 3.74 | -5.65 | -1.1 | 3.95 | -5.58 |
| 25 | 2.9 | 4.07 | -5.69 | 4.5 | 3.52 | -5.65 | 1.2 | 4.00 | -5.65 |
| 26 | 0.8 | 4.07 | -5.68 | 5.4 | 3.61 | -5.66 | -1.7 | 4.01 | -5.65 |
| 27 | 4.2 | 4.07 | -5.67 | -0.4 | 3.95 | -5.64 | -0.9 | 4.04 | -5.65 |
| 28 | -4.3 | 4.10 | -5.71 | 34.0 | 3.89 | -5.67 | 34.0 | 4.08 | -5.69 |
| 29 | 7.7 | 4.07 | -5.68 | 5.6 | 3.41 | -5.62 | 2.0 | 3.90 | -5.67 |
| 30 | -1.2 | 4.07 | -5.68 | 31.6 | 3.49 | -5.68 | 23.9 | 3.92 | -5.65 |
| 31 | -0.4 | 4.06 | -5.67 | 31.2 | 3.54 | -5.68 | 28.9 | 3.95 | -5.69 |
| 32 | 6.1 | 4.06 | -5.68 | 9.2 | 3.79 | -5.67 | 5.6 | 4.04 | -5.68 |
| 33 | -2.8 | 4.07 | -5.67 | 32.6 | 3.53 | -5.68 | 30.4 | 3.92 | -5.67 |
| 34 | 2.2 | 4.08 | -5.69 | 0.8 | 3.75 | -5.68 | -1.8 | 4.04 | -5.68 |
| 35 | 1.6 | 4.07 | -5.70 | 1.4 | 3.65 | -5.63 | -2.6 | 4.05 | -5.66 |
| 36 | 5.3 | 4.07 | -5.67 | 0.1 | 3.89 | -5.66 | -3.1 | 4.04 | -5.67 |
| 37 | 5.1 | 4.06 | -5.66 | **-0.9** | **3.90** | **-5.66** | **-6.0** | **4.06** | **-5.69** |
| MIN | -10.1 | 4.02 | -5.75 | -0.9 | 3.41 | -5.84 | -6.0 | 3.88 | -5.73 |
| MAX | 10.0 | 4.18 | -5.66 | 43.5 | 4.13 | -5.62 | 54.1 | 4.14 | -5.49 |
| AVERAGE | 1.0 | 4.08 | -5.69 | 16.8 | 3.77 | -5.68 | 13.9 | 4.02 | -5.66 |
| STD DEV | 4.8 | 0.03 | 0.02 | 16.5 | 0.19 | 0.05 | 18.2 | 0.06 | 0.04 |

**Table S10.** Empirical Lewis acidities (LA) of selected metal cations bonded to $O^{2-}$.[10] Note that lower LA corresponds to softer acids, while higher LA corresponds to harder acids. Additionally, carboxylates are generally classified as harder bases while phosphines are classified as softer bases.[11]

| Metal cation | LA |
|---|---|
| Sc(III) | 0.481 |
| Zn(II) | 0.405 |
| Ga(III) | 0.636 |
| Y(III) | 0.393 |
| Cd(II) | 0.320 |
| In(III) | 0.495 |

**Table S11.** Covalent radii of the elements studied in this work. The covalent radii (S) are used to identify bonds between atoms when generating the molecular graph for each structure. A molecular graph aids in identifying ligand binding modes and determining the local chemical environment around each In.

| elements | S (Å) |
|---|---|
| H | 0.31 |
| O | 0.66 |
| C | 0.73 |
| In | 1.42 |
| Zn | 1.22 |
| Sc | 1.70 |
| Ga | 1.22 |
| Y | 1.90 |
| Cd | 1.44 |
| P | 1.07 |

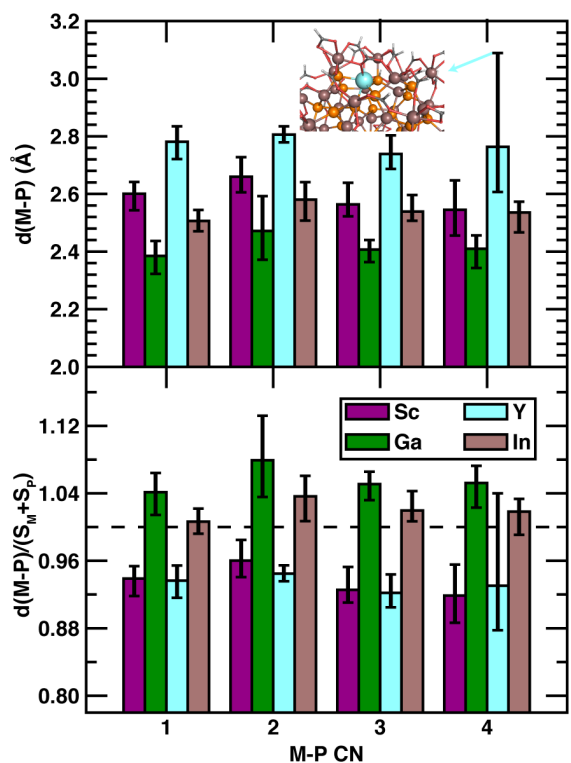

**Figure S8.** M-P bond distances (in Å, upper pane) and sum of covalent radii ($S$)-normalized M-P bond distances (lower pane) at each M-P coordination number (CN) across all relaxed M(III) doped (M=Ga,Sc,Y) MSC clusters and the relaxed undoped MSC cluster (M=In). Colored bar heights indicate the average over all relevant bond distances while error bars indicate minimum and maximum bond distances. A dashed horizontal line at 1.0 corresponds to where the M-P bond distance is the same as the sum of covalent radii. The inset in upper panel shows the Y-doped MSC at In index 21 having the greatest restructuring. Note that Sc(III) and Y(III) show lower normalized M-P bond distances at higher M-P CNs corresponding to compressed bonds.

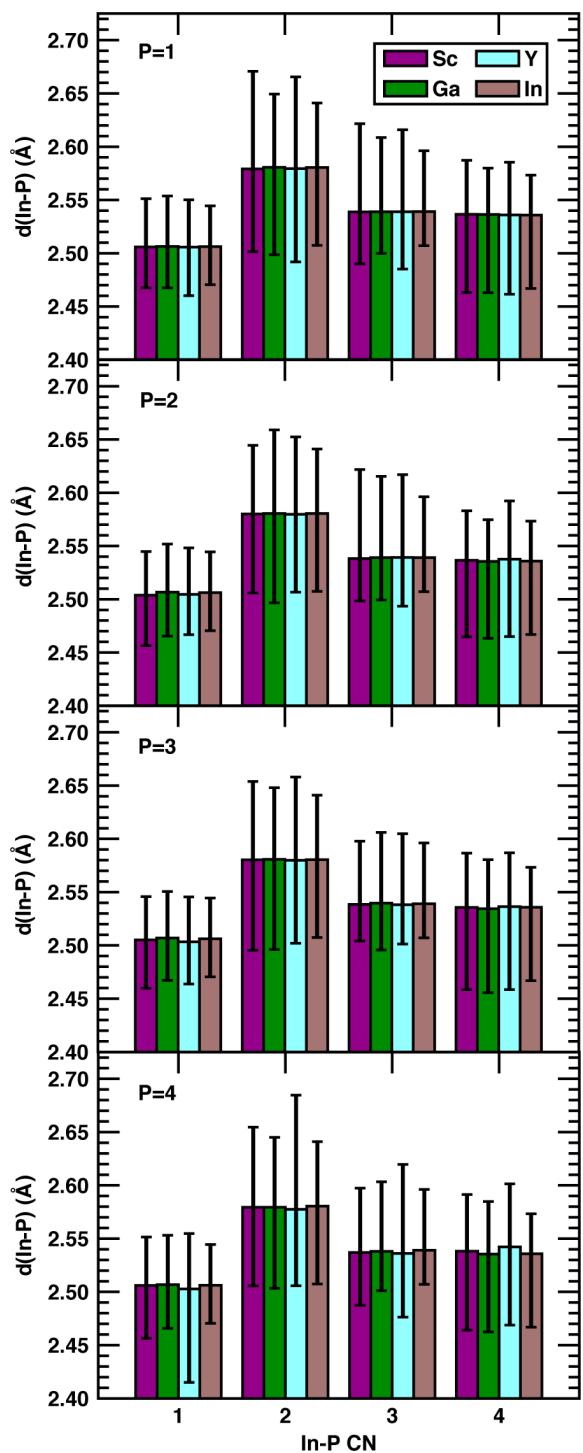

**Figure S9.** In-P bond distances (in Å) at each In-P coordination number (CN) across all relaxed M(III) doped (M = Ga, Sc, Y) MSC clusters and the relaxed undoped MSC cluster (M = In). Degree of P-coordination for the dopant metal increases from the top panel (P = 1) to the bottom panel (P = 4). Colored bar heights indicate the average over all relevant bond distances while error bars indicate minimum and maximum bond distances. We note that although variations in average bond lengths appear small, these are averaged over many times more bond distances than in Figure S8, and small shifts can correspond to larger strain distributed across the MSC.

Page S15

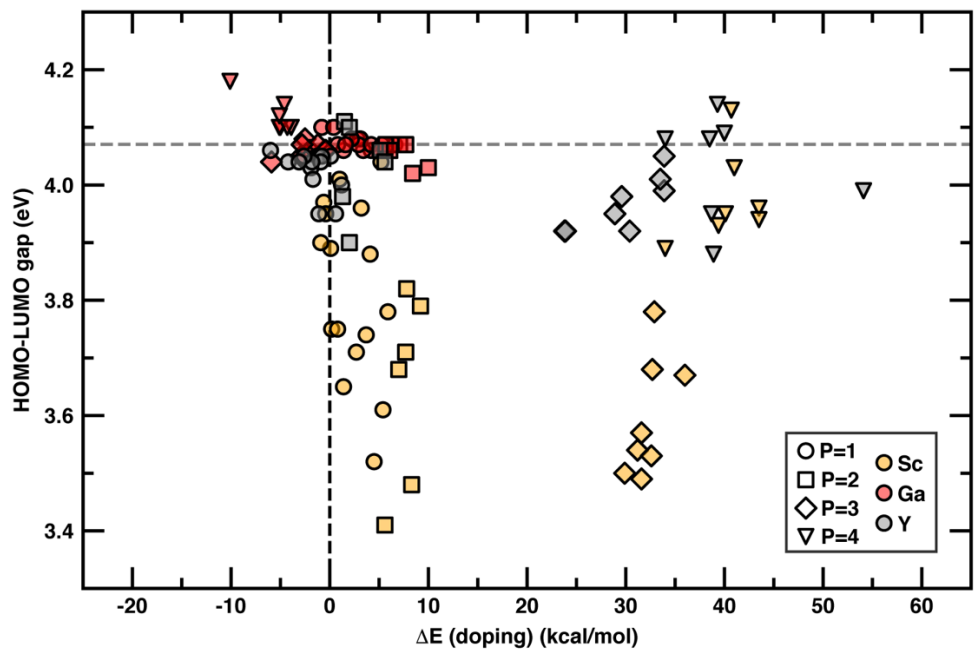

**Figure S10.** Correlation of MSC single M(III) (Ga=red, Sc=yellow, and Y=gray) ΔE (doping) in kcal/mol with HOMO-LUMO gap in eV. P-coordination number of the given site is distinguished by symbol with P=1 as circles, P=2 as squares, P=3 as diamonds, and P=4 as triangles. The vertical black line indicates where ΔE (doping)=0, while the horizontal dashed gray line indicates the HOMO-LUMO gap of the undoped MSC (4.07 eV). More favorable Ga doping to the upper left indicate slight increases in energy corresponding to blue shifts in low-energy UV-Vis absorption spectra, while primarily Y and Sc doping in the bottom right section indicate slight redshifts.

**Table S12.** Metal-P distance (Å) statistics split by degree of P coordination. M-P distances are taken from each geometry-relaxed structure.

| metal | P | N | AVERAGE | MIN | MAX | STD DEV |
|---|---|---|---|---|---|---|
| Sc | 1 | 16 | 2.60 | 2.54 | 2.64 | 0.03 |
|  | 2 | 12 | 2.66 | 2.61 | 2.73 | 0.03 |
|  | 3 | 24 | 2.56 | 2.52 | 2.64 | 0.03 |
|  | 4 | 28 | 2.55 | 2.46 | 2.65 | 0.04 |
| Zn | 1 | 38 | 2.41 | 2.34 | 2.49 | 0.03 |
|  | 2 | 20 | 2.45 | 2.39 | 2.53 | 0.03 |
|  | 3 | 24 | 2.43 | 2.38 | 2.47 | 0.02 |
|  | 4 | 0 |  |  |  |  |
| Ga | 1 | 16 | 2.38 | 2.32 | 2.44 | 0.03 |
|  | 2 | 12 | 2.47 | 2.37 | 2.59 | 0.07 |
|  | 3 | 24 | 2.41 | 2.36 | 2.44 | 0.02 |
|  | 4 | 28 | 2.41 | 2.34 | 2.46 | 0.03 |
| Y | 1 | 16 | 2.78 | 2.72 | 2.83 | 0.03 |
|  | 2 | 12 | 2.81 | 2.78 | 2.83 | 0.02 |
|  | 3 | 24 | 2.74 | 2.69 | 2.80 | 0.03 |
|  | 4 | 28 | 2.76 | 2.61 | 3.09 | 0.09 |
| Cd | 1 | 64 | 2.57 | 2.50 | 2.65 | 0.03 |
|  | 2 | 36 | 2.63 | 2.56 | 2.71 | 0.04 |
|  | 3 | 24 | 2.59 | 2.54 | 2.64 | 0.03 |
|  | 4 | 28 | 2.64 | 2.57 | 2.82 | 0.06 |
| In/MSC | 1 | 16 | 2.51 | 2.47 | 2.54 | 0.02 |
|  | 2 | 12 | 2.58 | 2.51 | 2.64 | 0.04 |
|  | 3 | 24 | 2.54 | 2.51 | 2.60 | 0.03 |
|  | 4 | 28 | 2.54 | 2.47 | 2.57 | 0.03 |

**Table S13.** MSC single M(II) (Cd and Zn) min, max, average, and standard deviation of ΔE (doping) in kcal/mol for each ligand binding mode.

| metal | mode | MIN | MAX | AVERAGE | STD DEV |
|---|---|---|---|---|---|
| Zn | ABF | -5.6 | 19.1 | 5.7 | 5.3 |
| Zn | CB | -11.6 | -2.2 | -6.2 | 2.9 |
| Zn | SB | 0.0 | 17.5 | 6.1 | 6.0 |
| Cd | ABF | -8.9 | 16.0 | 5.7 | 5.2 |
| Cd | CB | -9.8 | -3.1 | -6.3 | 2.1 |
| Cd | SB | -4.0 | 18.7 | 5.7 | 7.5 |

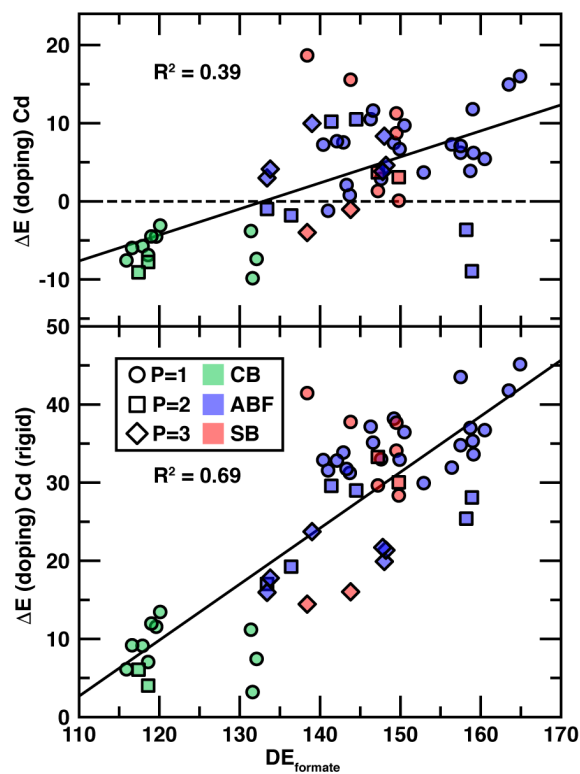

**Figure S11.** Plot of relaxed (upper) and rigid (lower) Cd ΔE (doping) in kcal/mol over the MSC sites versus the corresponding rigid formate ligand dissociation energy (DE$_{formate}$) in kcal/mol.[5] Black lines indicate the least squares regression fit, while a black dashed line indicates ΔE (doping) = 0 kcal/mol in the upper panel. Separation between ligand binding modes is overestimated in rigid Cd doping compared to relaxed Cd doping in addition to greater separation of doping energetics by binding mode.

**Table S14.** MSC single M(II) (Cd and Zn) min, max, average, and standard deviation of ΔE doping in kcal/mol for each P-coordination number.

| metal | P | MIN | MAX | AVERAGE | STD DEV |
|---|---|---|---|---|---|
| Zn | 1 | -11.6 | 19.1 | 3.2 | 7.3 |
| Zn | 2 | -9.1 | 13.5 | 1.2 | 7.7 |
| Zn | 3 | 0.0 | 11.0 | 5.5 | 3.7 |
| Cd | 1 | -9.8 | 18.7 | 4.0 | 7.5 |
| Cd | 2 | -9.1 | 10.5 | -0.5 | 7.3 |
| Cd | 3 | -4.0 | 10.0 | 3.6 | 4.5 |

**Table S15.** MSC single M(II) (Zn and Cd) doping energetics including ΔE (doping) in kcal/mol (ΔE, in kcal/mol) and HOMO-LUMO gap (gap, in eV) with reference In indices that are replaced by M(II) and the ligand index removed upon doping. The ligand binding modes and M-P CN are included for reference. The min, max, average, and standard deviation values are shown at the bottom. The HOMO-LUMO gap of the undoped MSC is 4.07 eV.

| In index | ligand index | mode | M-P CN | Zn ΔE | Zn gap | Cd ΔE | Cd gap | \|ΔE$_{Zn}$ - ΔE$_{Cd}$\| |
|---|---|---|---|---|---|---|---|---|
| 1 | 2 | ABF | 1 | 4.9 | 4.03 | 7.3 | 4.01 | 2.34 |
| 1 | 6 | ABF | 1 | 9.2 | 3.92 | 11.8 | 3.91 | 2.61 |
| 2 | 1 | ABF | 1 | 5.5 | 4.01 | 6.2 | 4.02 | 3.32 |
| 2 | 7 | CB | 1 | -2.9 | 4.05 | -5.7 | 4.04 | 0.59 |
| 3 | 10 | CB | 1 | -5.3 | 4.06 | -4.5 | 4.04 | 3.57 |
| 3 | 3 | ABF | 1 | 8.7 | 4.03 | 6.2 | 4.04 | 4.70 |
| 4 | 4 | CB | 2 | -7.8 | 4.03 | -9.2 | 4.00 | 1.58 |
| 5 | 19 | SB | 1 | 17.5 | 3.57 | 15.6 | 3.59 | 1.01 |
| 5 | 5 | ABF | 1 | 6.7 | 4.01 | 6.7 | 3.99 | 3.98 |
| 5 | 8 | CB | 1 | **-11.6** | **4.02** | **-9.8** | **4.01** | 3.18 |
| 6 | 9 | CB | 1 | -3.5 | 4.04 | -6.0 | 4.01 | 1.80 |
| 6 | 20 | ABF | 1 | 7.3 | 4.00 | 5.4 | 4.03 | 1.19 |
| 7 | 11 | ABF | 3 | 4.8 | 3.98 | 3.8 | 3.90 | 2.47 |
| 8 | 13 | ABF | 2 | -5.6 | 4.07 | -9.0 | 4.05 | 0.70 |
| 8 | 16 | ABF | 2 | 3.3 | 4.06 | -1.0 | 4.03 | 2.85 |
| 11 | 24 | SB | 2 | 3.7 | 4.05 | 3.1 | 4.04 | 1.56 |
| 11 | 12 | ABF | 2 | 13.5 | 3.98 | 10.2 | 3.98 | 2.48 |
| 12 | 14 | ABF | 1 | 3.7 | 4.03 | 7.3 | 3.97 | 3.38 |
| 12 | 29 | SB | 1 | 2.9 | 4.02 | 1.3 | 4.01 | 4.01 |
| 12 | 15 | ABF | 1 | 2.9 | 4.00 | 7.6 | 4.00 | 1.84 |
| 13 | 21 | ABF | 3 | 11.0 | 3.97 | 10.0 | 3.90 | 2.35 |
| 14 | 22 | SB | 3 | 0.0 | 4.02 | -4.0 | 3.97 | 3.77 |
| 15 | 19 | SB | 3 | 2.1 | 3.98 | -1.0 | 4.01 | 1.14 |
| 17 | 17 | ABF | 2 | 0.0 | 4.04 | -1.8 | 3.98 | 4.12 |
| 17 | 25 | ABF | 2 | -2.5 | 4.07 | -3.6 | 4.05 | 1.10 |
| 18 | 18 | ABF | 3 | 5.5 | 4.04 | 4.0 | 3.94 | 3.22 |
| 20 | 27 | CB | 1 | -2.2 | 4.05 | -3.8 | 4.04 | 1.85 |
| 20 | 36 | ABF | 1 | 5.0 | 4.02 | 7.5 | 4.02 | 2.51 |
| 23 | 38 | SB | 1 | 7.3 | 3.99 | 11.3 | 3.96 | 2.74 |
| 23 | 26 | CB | 1 | -7.9 | 4.07 | -4.5 | 4.05 | 2.29 |
| 24 | 40 | ABF | 1 | 3.9 | 3.96 | 7.7 | 3.95 | 0.32 |
| 24 | 24 | SB | 1 | 1.9 | 4.02 | 0.1 | 4.04 | 0.92 |
| 24 | 32 | ABF | 1 | 3.1 | 4.01 | 0.8 | 3.99 | 0.95 |
| 25 | 28 | ABF | 1 | 1.0 | 4.10 | 2.1 | 4.06 | 0.79 |
| 25 | 33 | ABF | 1 | 19.1 | 3.98 | 15.0 | 3.96 | 2.50 |
| 25 | 39 | ABF | 1 | 9.4 | 3.95 | 10.5 | 3.94 | 1.12 |
| 26 | 30 | ABF | 1 | 1.9 | 4.08 | 3.7 | 4.07 | 2.06 |
| 26 | 34 | ABF | 1 | 18.5 | 3.98 | 16.0 | 3.95 | 1.38 |
| 26 | 23 | ABF | 1 | 8.4 | 3.98 | 11.6 | 3.97 | 0.54 |
| 27 | 22 | SB | 1 | 15.9 | 3.54 | 18.7 | 3.55 | 3.95 |
| 27 | 41 | ABF | 1 | 4.2 | 4.01 | 3.9 | 3.98 | 2.44 |
| 27 | 31 | CB | 1 | -9.2 | 4.04 | -6.9 | 4.02 | 8.74 |
| 29 | 29 | SB | 2 | 4.6 | 4.02 | 3.7 | 4.01 | 1.74 |
| 29 | 37 | ABF | 2 | 11.5 | 4.00 | 10.5 | 3.99 | 0.78 |
| 30 | 45 | ABF | 3 | 5.3 | 4.03 | 4.1 | 3.90 | 3.14 |
| 31 | 44 | ABF | 3 | 10.4 | 3.98 | 8.3 | 3.90 | 0.59 |
| 32 | 43 | CB | 2 | -9.2 | 4.04 | -7.8 | 4.00 | 2.84 |
| 33 | 48 | ABF | 3 | 5.2 | 3.98 | 4.6 | 3.89 | 1.26 |
| 34 | 47 | CB | 1 | -5.5 | 4.07 | -3.1 | 4.05 | 1.97 |
| 34 | 38 | SB | 1 | 4.8 | 4.01 | 8.7 | 3.98 | 0.05 |
| 35 | 46 | ABF | 1 | -2.9 | 4.04 | -1.2 | 4.02 | 1.80 |
| 35 | 35 | ABF | 1 | 1.0 | 4.01 | 9.7 | 4.00 | 1.86 |
| 36 | 42 | ABF | 1 | 6.3 | 3.98 | 7.1 | 3.98 | 2.45 |
| 36 | 50 | CB | 1 | -4.4 | 4.04 | -7.6 | 4.02 | 0.97 |
| 37 | 51 | CB | 1 | -4.5 | 4.05 | -7.4 | 4.01 | 3.33 |
| 37 | 49 | ABF | 1 | 2.3 | 4.02 | 2.9 | 4.02 | 4.27 |
| | | | MIN | -11.6 | 3.54 | -9.8 | 3.55 | 0.05 |
| | | | MAX | 19.1 | 4.10 | 18.7 | 4.07 | 8.73 |
| | | | AVERAGE | 3.2 | 4.00 | 3.2 | 3.98 | 2.27 |
| | | | STD DEV | 7.0 | 0.09 | 7.2 | 0.09 | 1.45 |

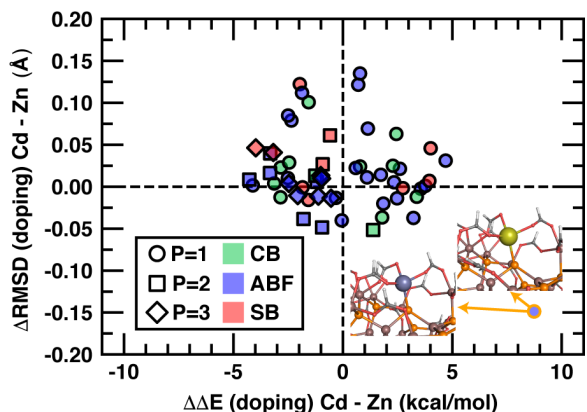

**Figure S12.** Plot of the difference in root mean square displacement of all atoms in the doped MSC compared to the undoped MSC (ΔRMSD, in Å) versus the difference in doping energy (ΔΔE (doping), in kcal/mol) of Cd(II) and Zn(II) for the MSC sites. Black dashed lines indicate ΔΔE (doping) = 0 kcal/mol and ΔRMSD = 0 Å. The inset highlights the outlier point, where the Zn(II) dopant (purple-gray) causes restructuring in the MSC compared to the equivalent Cd(II) dopant (yellow).

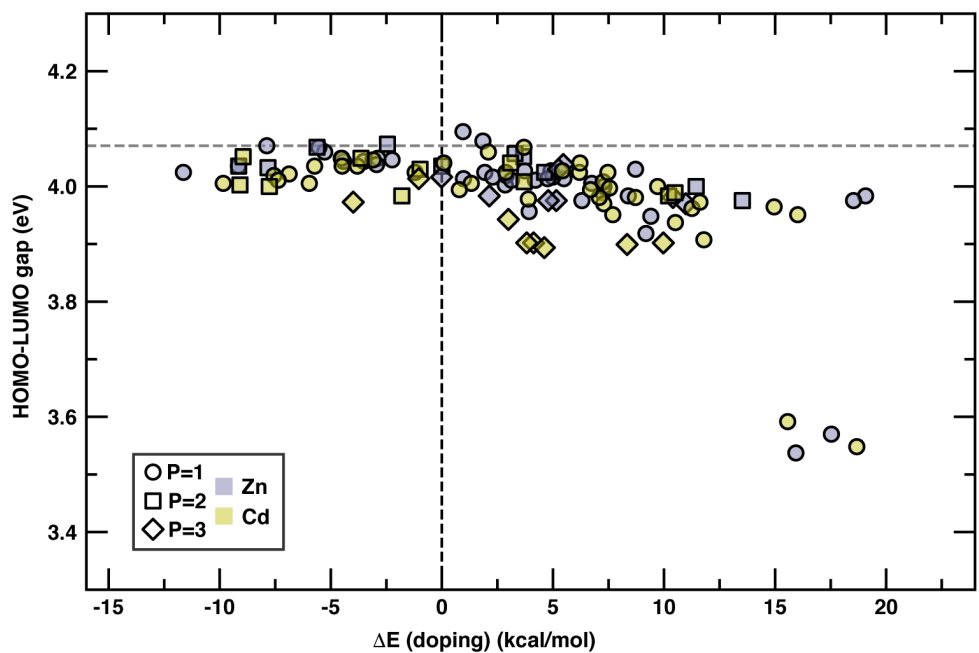

**Figure S13.** Plot of MSC single M(II) (Zn in purple-gray and Cd in yellow) ΔE (doping) in kcal/mol with HOMO-LUMO gap in eV. The P-coordination number of the given site is distinguished by symbol with P=1 as circles, P=2 as squares, and P=3 as diamonds. The vertical black dashed line indicates where ΔE (doping) = 0 kcal/mol, while the horizontal dashed gray line indicates the HOMO-LUMO gap of the undoped MSC (i.e., 4.07 eV). More favorable Cd and Zn doping to the upper left is associated with slight decreases in HOMO-LUMO gap corresponding to redshifts in low-energy UV-Vis absorption spectra.

**Table S16.** MSC multiple dopant ΔE (doping) in kcal/mol and In replaced and ligand removed indices for lowest ΔE (doping) configurations of Cd(II), Zn(II), and Ga(III). The predicted (pred ΔE) energetics for the indium and ligands indices corresponding to the sum of the single dopant site energetics are reported in kcal/mol. The correspondence between the actual lowest ΔE configuration and the lowest predicted (i.e., from single dopants) ΔE configuration is indicated with True (T) or False (F). All lowest ΔE configurations are either the lowest or second lowest pred ΔE. The number of ligands removed that are shared between two adjacent doped sites are also tabulated ($n_{bridging}$). The average M-P coordination number (CN) of the dopants are included ($P_{avg}$). The change in ΔE from the prior number of dopants (ΔΔE) is reported in kcal/mol. HOMO-LUMO gap of the structures (gap) are reported in eV.

| metal | $n_{dopants}$ | indium indices | ligand indices | $n_{bridging}$ | $P_{avg}$ | Low ΔE? | gap | ΔE | ΔΔE | pred ΔE |
|---|---|---|---|---|---|---|---|---|---|---|
| Cd | 1 | 5 | 8 | 0 | 1.0 | T | 4.01 | -9.8 | -9.8 | -9.8 |
|  | 3 | 32,4,5 | 43,4,8 | 0 | 1.7 | F | 3.92 | -28.1 | -16.9 | -26.7 |
|  | 5 | 32,37,4,5,8 | 43,51,4,8,13 | 1 | 1.6 | F | 3.78 | -35.7 | -16.3 | -43.0 |
|  | 7 | 27,32,36,37,4,5,8 | 31,43,50,51,4,8, 13 | 1 | 1.4 | T | 3.75 | -49.3 | -14.4 | -57.4 |
|  | 9 | 23,27,32,36,37,4,5,6,8 | 26,31,43,50,51,4,8,9,13 | 1 | 1.3 | F | 3.71 | -60.2 | -10.5 | -67.9 |
|  | 11 | 23,27,2,32,36,37,3,4,5, 6,8 | 26,31,7,43,50,51,10,4,8, 9,13 | 1 | 1.3 | T | 3.58 | -64.5 | -10.2 | -78.1 |
|  | 13 | 20,23,27,2,32,36,37,3, 4,5,6,14,8 | 27,26,31,7,43,50,51,10, 4,8,9,22,13 | 2 | 1.4 | T | 3.46 | -68.6 | -7.8 | -85.9 |
|  | 15 | 20,23,27,2,32,34,36, 37,3,4,5,6,14,17,8 | 27,26,31,7,43,47,50,51 ,10,4,8,9,22, 25,13 | 3 | 1.4 | T | 3.48 | -68.7 | -6.7 | -92.7 |
| Zn | 1 | 5 | 8 | 0 | 1.0 | T | 4.02 | -11.6 | -11.6 | -11.6 |
|  | 3 | 27,32,5 | 31,43,8 | 0 | 1.3 | T | 3.96 | -30.3 | -18.3 | -29.9 |
|  | 5 | 23,27,32,4,5 | 26,31,43,4,8 | 0 | 1.4 | T | 3.96 | -46.5 | -15.7 | -45.7 |
|  | 7 | 23,27,32,34,4,5,8 | 26,31,43,47,4,8,13 | 1 | 1.4 | T | 3.97 | -53.1 | -11.2 | -56.8 |
|  | 9 | 23,27,32,34,36,3,4,5 ,8 | 26,31,43,47,50,10,4,8, 13 | 1 | 1.3 | F | 3.95 | -55.3 | -9.7 | -66.5 |
|  | 11 | 23,27,32,34,36,37,3, 4, 5,6,8 | 26,31,43,47,50,51,10,4 ,8, 9,13 | 1 | 1.3 | T | 3.89 | -56.1 | -8.0 | -74.5 |
|  | 13 | 17,23,27,32,34,35,3 6, 37, 3,4,5,6,8 | 25,26,31,43,47,46,50,5 1, 10,4,8,9,13 | 3 | 1.3 | F | 3.83 | -56.4 | -5.4 | -79.9 |
|  | 15 | 17,20,23,27,2,32,34, 35,36,37,3,4,5,6,8 | 25,27,26,31,7,43, 47,46,50,51,10, 4,8,9,13 | 3 | 1.3 | T | 3.77 | -63.6 | -5.1 | -85.1 |
| Ga | 1 | 19 |  | 0 | 4.0 | T | 4.18 | -10.1 | -10.1 | -10.1 |
|  | 3 | 15,19,21 |  | 0 | 3.7 | F | 4.21 | -20.2 | -11.0 | -21.1 |
|  | 5 | 10,15,19,21,22 |  | 0 | 3.8 | T | 4.32 | -31.2 | -10.2 | -31.3 |
|  | 7 | 10,15,16,19,21,22,9 |  | 0 | 3.9 | F | 4.43 | -39.1 | -8.5 | -39.7 |
|  | 9 | 10,14,15,16,19,21,2 2, 28,9 |  | 0 | 3.8 | T | 4.46 | -48.2 | -7.3 | -47.0 |
|  | 11 | 10,14,15,16,19,21,2 2,28, 33,7,9 |  | 0 | 3.6 | T | 4.49 | -54.8 | -5.3 | -52.3 |
|  | 13 | 10,13,14,15,16,19,2 1,22, 28,30,33,7,9 |  | 0 | 3.5 | T | 4.51 | -58.7 | -3.2 | -55.4 |
|  | 15 | 10,12,13,14,15,16,1 8,19, 21,22,28,30,33,7,9 |  | 0 | 3.3 | T | 4.52 | -63.3 | -1.3 | -56.7 |

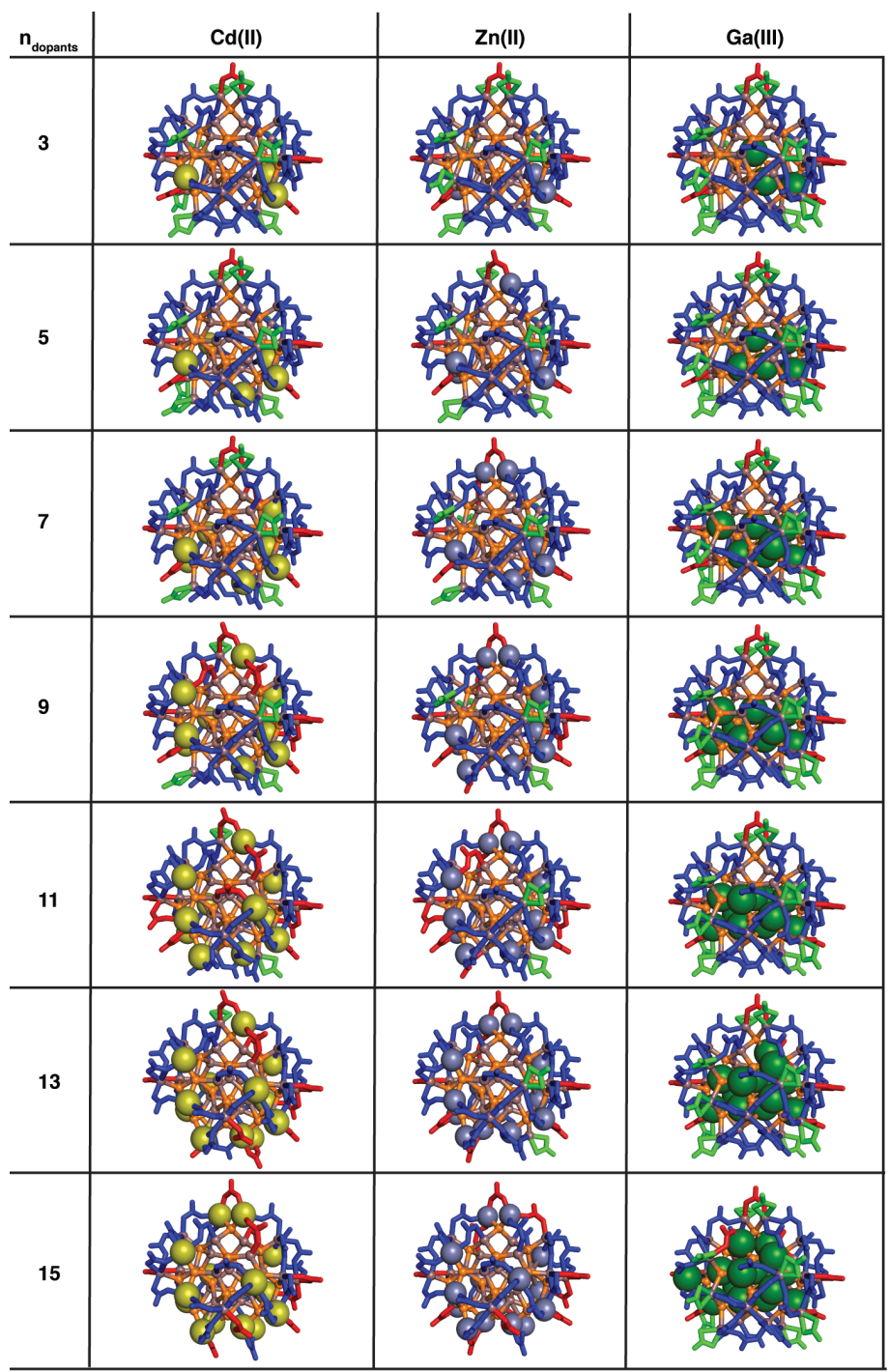

**Figure S14.** Structures for the minimum-energy configuration of each higher dopant concentration ($n_{dopants}$) for Cd(II) (yellow, left), Zn(II) (gray-purple, middle), and Ga(III) (dark green, right). For each structure the ligand binding modes are colored with asymmetric bridging ligands in blue, symmetric bridging ligands in red, and chelating bidentate ligands in green. Note that for Zn(II) and Cd(II) largely chelating bidentate ligands are replaced as the surface becomes saturated with dopants. In the Ga(III)-doped cases, ligands are not removed with increasing dopant concentration, and the lowest-energy configurations consist entirely of interior doping sites up to $n_{dopants} = 15$.

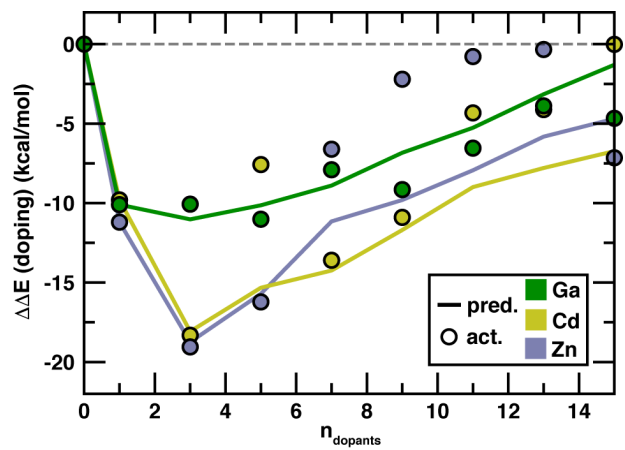

**Figure S15.** Plot of the change in the ΔE (doping) relative to the ΔE (doping) of the next lowest $n_{dopants}$ (ΔΔE (doping) in kcal/mol) for lowest-energy configurations sampled at each $n_{dopants}$ versus number of dopants. $n_{dopants}$ represents a concentration doped in the MSC, where there is a total of 37 In-metal sites for cation exchange, meaning 15 dopants represents a concentration of 41% in the MSC. Predicted (pred.) ΔE (doping) energetics represents the sum of the equivalent single site ΔE (doping) energetics, while actual (act.) represent the calculated values with multiple dopants.

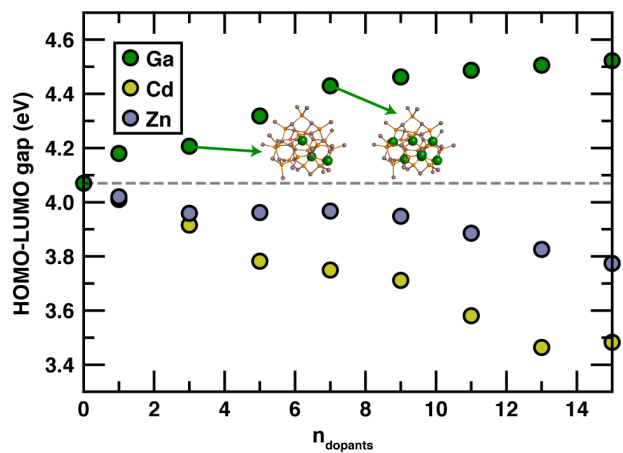

**Figure S16.** Plot of doped MSC HOMO-LUMO gap in eV versus $n_{dopants}$ for Zn(II), Cd(II), and Ga(III). The horizontal dashed gray line indicates the HOMO-LUMO gap of the undoped MSC (i.e., 4.07 eV). All dopants show equivalent red/blue shifts in low-energy UV-vis absorption spectra as the single dopants. Cd(II) shows the largest shifts in HOMO-LUMO gap and is shows notable decreasing gap with increasing concentration. Zn(II) results relatively minor shifts in gap, but does show decreasing gap with increasing concentrations. Ga(III) shows larger shifts in gap from 3 to 7 Ga(III) doped as bulk In(III) is replaced with Ga(III) as shown in the insets.

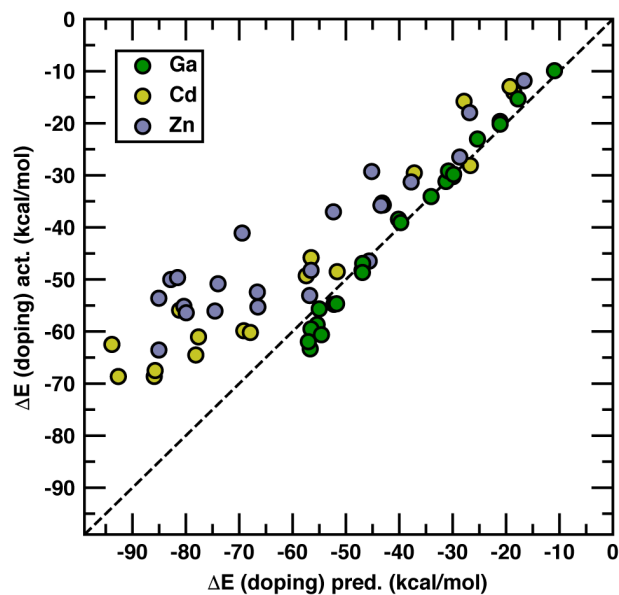

**Figure S17.** Parity plot between actual (act.) and predicted (pred.) ΔE (doping) for Zn(II), Cd(II), and Ga(III) across all higher doping configurations (all in kcal/mol). The black dashed line indicates ΔE (doping) pred. = ΔE (doping) act. Ga(II) shows the least deviations between pred. and act. with mean absolute error (MAE) of 2.06 kcal/mol followed by Cd(II) (MAE =12.2 kcal/mol) followed by Zn(II) (MAE=15.3 kcal/mol).